\documentclass[12pt]{article}
\usepackage{amsmath}
\usepackage[utf8]{inputenc}
\usepackage[a4paper,left=2cm,right=2cm,top=2cm,bottom=2cm]{geometry}
\def\Lag{{\mathcal L}{}}
\def\Sag{{\mathcal S}{}}
\def\Eag{{\mathcal E}{}}
\def\Kag{{\mathcal K}{}}
\def\Mag{{\mathcal M}{}}

\def\tref{{}^{0}h}

\def\tref{{}^{0}h}

\def\ombol{{\stackrel{\circ}{\omega}}{}}

\def\Rbol{{\stackrel{\circ}{R}}{}}
\def\Gammabol{{\stackrel{\circ}{\Gamma}}{}}
\def\nablabol{{\stackrel{\circ}{\nabla}}{}}

\usepackage{setspace}

\usepackage[colorlinks=true,urlcolor=blue]{hyperref}
\usepackage{etex}
\usepackage{amsmath,amsfonts,amsbsy,amssymb,amscd,latexsym}
\usepackage{times}
\usepackage[sort&compress,numbers]{natbib}
\usepackage[affil-it]{authblk}

\usepackage{xcolor}

\begin{document}
\title{ \bf  Teleparallel Renormalization of Gravitational Action}
\author{Martin Kr\v{s}\v{s}\'ak\thanks{Electronic address: \texttt{martin.krssak@fmph.uniba.sk, martin.krssak@gmail.com}}}

\affil{Department of Theoretical Physics, Faculty of Mathematics, Physics and Informatics, Comenius University in Bratislava, 84248, Slovak Republic}
\affil{Department of Astronomy,  School of Physical Sciences, University of	Science and Technology of China, Hefei, Anhui 230026, China}

\author{Michal Stano\thanks{Electronic address: \texttt{stano@math.cas.cz, stano.michal.ml@gmail.com}}}
\affil{Institute of Mathematics of the Czech Academy of Sciences,\\
\v{Z}itn\'a 25, 115 67 Prague 1, Czech Republic}
\affil{Institute of Theoretical Physics, Faculty of Mathematics and Physics, Charles University, V Holešovičkách 2, 180 00 Prague 8, Czech Republic}

\date{\today}
\maketitle

\begin{abstract}
We study the gravitational action in the teleparallel equivalent of general relativity, where the spin connection contributes to the action through a surface term and, when chosen appropriately, renormalizes the action by effectively playing the role of a counterterm. We resolve several apparent ambiguities in this procedure using the Schwarzschild spacetime as an example. We show that the apparent discrepancy between bulk and quasilocal actions originates from contributions associated with spacetime singularities, and that the quasilocal action agrees with the standard results obtained in general relativity. We further clarify the role of the tetrad choice in determining the corresponding spin connection, showing that the proper diagonal and canonical frames give the same Euclidean action, while only the proper diagonal frame reproduces the expected Lorentzian action on the Wheeler--DeWitt patch. Finally, we establish the uniqueness of the renormalization scheme and show that imposing an appropriate fall-off behavior of the torsion tensor removes the remaining ambiguity and fixes the Euclidean action uniquely.
\end{abstract}

\newpage
\section{Introduction\label{secintro}}
The finite Euclidean action solutions of general relativity play an important role in the path integral approach to quantum gravity \cite{Hawking:1978jz,Hawking:1978jn,Gibbons:2002du}, black hole thermodynamics \cite{Gibbons:1976ue,York:1986it}, and holography \cite{Witten:1998qj}. Their construction is complicated by the fact that the presence of second derivatives of the metric in the Einstein-Hilbert action requires the addition of the Gibbons-Hawking-York (GHY) boundary term, which generally diverges and  needs to be renormalized by an appropriate counterterm \cite{York:1972sj,Gibbons:1976ue}. The most well-known counterterm is obtained using the original Gibbons-Hawking ``background subtraction" method \cite{Gibbons:1976ue, Hawking:1995fd}, but several alternative counterterms have been proposed since then \cite{Henningson:1998gx,Balasubramanian:1999re,Emparan:1999pm,Bianchi:2001kw, Mann:2005yr,Olea:2005gb,Anastasiou:2020zwc}.

An alternative approach 
is to consider the problem not within standard general relativity, but rather in its teleparallel formulation, where gravity is described using teleparallel geometry instead of the usual Riemannian one. This formulation is known as the teleparallel equivalent of general relativity \cite{Aldrovandi:2013wha,Maluf:2013gaa,Krssak:2018ywd}, or simply teleparallel gravity, and has a long history dating back to Einstein's attempt to construct a unified theory \cite{Einstein1928b,Einstein1929a}. Although teleparallelism as a unified theory was ultimately unsuccessful \cite{Sauer:2004hj,Goenner:2004se,Krssak:2024xeh}, it has found applications in the study of conserved quantities \cite{Moller1961,Pellegrini1963,Moller1966,Moller1978}, the gauge aspects of gravity \cite{Cho:1975dh,Fontanini:2018krt,Pereira:2019woq,LeDelliou:2019esi,Brezina:2025dbc}, and more recently, in various extended models of gravity aimed at addressing problems in cosmology \cite{Ferraro:2006jd,Ferraro:2008ey,Linder:2010py,Cai:2015emx}.

An interesting feature of the teleparallel formulation is that the teleparallel gravitational action does not contain second derivatives of the dynamical variables and hence does not require the addition of GHY-like boundary terms. However, teleparallel quantities--including the action--depend not only on the tetrad field but also on the non-dynamical pure-gauge teleparallel connection, which is not determined by the field equations. It turns out that, when the teleparallel connection is chosen appropriately to match the tetrad, its contribution to the action is a surface term that renders the action finite \cite{Krssak:2015rqa, Krssak:2015lba,Krssak:2024kva}. This can be viewed as an alternative method of renormalizing the gravitational action, where the problem of finding the appropriate counterterm in general relativity is replaced by the problem of determining the spin connection that matches the tetrad. We will refer to this as the \textit{teleparallel renormalization} of the gravitational action.

Here it is useful to clarify our choice of terminology and note that the term \textit{teleparallel regularization} would be equally appropriate, since the role of the spin connection  is to render the gravitational action finite by removing divergences. However, since determining the spin connection is also equivalent to finding a suitable counterterm \cite{Krssak:2015lba,Krssak:2024kva}, the procedure bears a close resemblance to holographic renormalization \cite{Bianchi:2001kw,Skenderis:2002wp}. The choice between the terms \textit{regularization} and \textit{renormalization} therefore reflects which aspect of the procedure one wishes to emphasize: the removal of divergences and the resulting finite action, or the identification of a counterterm that achieves this. We adopt the latter terminology, in accordance with our previous works \cite{Krssak:2015rqa,Krssak:2015lba,Krssak:2024kva}\footnote{However, we would like to stress that this is not the UV renormalization in the standard QFT sense, but rather in the  spirit of  holographic renormalization.}. 

A natural question is whether teleparallel renormalization agrees with the standard prescription of general relativity and whether it leads to unique predictions, with different answers having been obtained in the literature.  In the original paper \cite{Krssak:2015rqa}, where the teleparallel spin connection was determined for the standard diagonal tetrad for the Schwarzschild solution, it was shown that the gravitational action obtained as a volume integral leads to twice the value found in general relativity\footnote{A similar result was also found in symmetric teleparallel gravity \cite{BeltranJimenez:2018vdo}.}. This was followed by the work of Oshita and Wu \cite{Oshita:2017nhn}, in which the teleparallel action was transformed into a surface term using the equivalence with the Einstein-Hilbert action and Stokes’ theorem. After subtracting the contribution from a reference spacetime,  exact agreement with the result of Gibbons and Hawking was found. On the other hand, evaluating the Lorentzian action over the  Wheeler--DeWitt patch was found to reproduce the general-relativistic result even when the action was evaluated as a volume integral \cite{Krssak:2023nrw}.
Further ambiguity arose from considering alternative choices of frame rather than starting with the  diagonal tetrad. In particular, the so-called canonical frame, defined by the vanishing of the gravitational energy-momentum pseudotensor, was proposed in \cite{BeltranJimenez:2019bnx} and argued to be unique and reproduce the correct result \cite{Koivisto:2022nar,Gomes:2022vrc,Gomes:2023hyk,BeltranJimenez:2024ufa}. More recently, free-falling frames were considered, for which the Euclidean action was found to be three times the Gibbons--Hawking result \cite{Fiorini:2023axr}, despite being evaluated as a surface integral. See also the relevant discussions  in \cite{Stano:2025eje,Kuntz:2024qgs}.

The aim of this paper is to understand and resolve these seemingly conflicting results and to establish teleparallel renormalization as a feasible alternative to the methods existing in general relativity. We  address several possible sources of the apparent ambiguities, starting with the distinction between the bulk action obtained by integrating the Lagrangian over the spacetime volume, and the quasilocal action obtained by rewriting the bulk action as a surface integral. We show that the use of the quasilocal action is the key to recovering agreement with the standard methods of general relativity, since the quasilocal action includes the contribution due to the central singularity, while the bulk action does not. We illustrate this on analogy with Gauss's law for the point sources in electrostatics and argue that the GHY action in general relativity includes these contributions, and hence we should include them in our teleparallel calculations as well.

Understanding the difference between the bulk and quasilocal actions then allows us to address an ambiguity in determining the teleparallel spin connection associated with a given tetrad. We refer to the resulting pair of a tetrad and its corresponding spin connection as the \textit{proper frame}. Different choices of the starting tetrad can lead to different proper frames. We consider here the proper diagonal and canonical frames and show that, for the Euclidean action, both frames yield the same result as in standard general relativity. 
However, for the Lorentzian action on the Wheeler-de Witt patch, which plays an important role in the calculations of holographic complexity \cite{Brown:2015bva,Brown:2015lvg}, only the proper diagonal frame leads to the correct results. 

Finally, the possibility of the existence of multiple frames with different results for the value of the quasilocal action then leads us to address the uniqueness of the teleparallel renormalization scheme in general. We show that the teleparallel spin connection is determined  only up to some residual local Lorentz transformations, which can  be classified by their asymptotic behavior. 
Transformations that decay sufficiently rapidly, corresponding to boosts whose rapidity falls off faster than $1/\sqrt{r}$, leave the action unchanged and therefore represent symmetries of the teleparallel renormalization procedure.
We further show that boosts with rapidity falling off precisely as $1/\sqrt{r}$ shift the action by a constant and would therefore introduce an ambiguity in the Euclidean action. We then argue that requiring all components of the torsion tensor, rather than only its vectorial part, to fall off as $\mathcal{O}(r^{-2})$ naturally excludes these transformations and leads to a unique value of the Euclidean action.

Let us clarify that teleparallel gravity can be formulated in two different ways, depending on how the teleparallel spin connection is treated. Throughout this paper, we use the covariant formulation, in which the fundamental variables are the tetrad and the pure-gauge teleparallel spin connection \cite{Obukhov:2002tm,Lucas:2009nq,Aldrovandi:2013wha,Krssak:2018ywd}. Due to the pure-gauge nature of the spin connection, one can choose the Weitzenb\"ock gauge, in which the spin connection vanishes, and thereby formulate the theory entirely in terms of the tetrad. This is known as the non-covariant, or pure-tetrad, formulation, which corresponds to Einstein's original approach to teleparallelism \cite{Einstein1928b,Einstein1929a} and is closely related to Møller's theory\cite{Moller1961}. Although the covariant formulation has the advantage of manifest local Lorentz invariance and avoids conceptual issues of the non-covariant formulation, such as identifying the tensorial torsion with the non-tensorial coefficients of anholonomy \cite{Aldrovandi:2013wha,Krssak:2024xeh}, the two formulations are effectively equivalent in practical calculations. This correspondence allows results obtained in the non-covariant formulation to be related to the teleparallel renormalization considered here, including the construction of proper tetrads \cite{Maluf:1995re,Maluf:1996kx}, the regularization of energy-momentum \cite{Obukhov:2006sk}, and black-hole thermodynamics \cite{Oshita:2017nhn,Fiorini:2023axr}. For a further discussion of the differences between the two formulations, see \cite{Krssak:2024xeh,Golovnev:2017dox,Maluf:2018coz,Golovnev:2023yla}.

\textit{Notation:} We follow the notation of \cite{Krssak:2024kva}. Geometric quantities with ``$\circ$" above them correspond to the Riemannian connection, while ``bare" quantities are defined with respect to the teleparallel connection. Latin indices denote tangent space coordinates, whereas Greek indices represent spacetime coordinates. We adopt the mostly-plus convention, $\eta=\text{diag}(-1,1,1,1)$, in the Lorentzian case. When working with specific components, we distinguish tangent-space indices using a hat. For example, in  $\omega^{\hat{1}}{}_{\hat{2}t}$, the first two indices refer to tangent-space components, while $t$ represents a spacetime component

\section{Geometric Formulations of General Relativity \label{chap2}}
The standard metric formulation of general relativity is based on the Riemannian geometry, where  the fundamental variable is  the metric tensor $g_{\mu\nu}$ defining uniquely the Riemannian linear  connection $\Gammabol^\rho{}_{\nu\mu}$, from which we obtain
the  curvature tensor $\Rbol^\rho{}_{\sigma\mu\nu}$, the Ricci tensor $\Rbol_{\mu\nu}=\Rbol^\rho{}_{\mu\sigma\nu}$ and the scalar curvature $\Rbol=g^{\mu\nu}\Rbol_{\mu\nu}$ \cite{Fecko:2006zy}.

The full gravitational action of general relativity is then given by the Einstein-Hilbert term taken as a linear function of the scalar curvature, the GHY boundary
term required by the well-posedness of the variational principle, and the counterterm \cite{York:1972sj,Gibbons:1976ue}
\begin{equation}\label{actiongr}
	\Sag_\text{GR}=\Sag_{\text{EH}}+\Sag_{\text{GHY}}+\Sag_{\text{counter}}=\frac{1}{2\kappa}\int_{\mathcal{M}}^{} \sqrt{-g} \Rbol +\frac{1}{\kappa}\int_{\partial \mathcal{M}}^{}  \sqrt{-\gamma} \left(\Kag - \Kag_0\right),
\end{equation}
where $\kappa=8\pi$ in natural units, $\gamma$ is the determinant of the boundary metric, and  $\Kag$ is its extrinsic curvature. The boundary is taken here to be a surface of some constant $r$-coordinate, which is then sent to infinity. We write the counterterm for the original Gibbons-Hawking boundary subtraction method \cite{Gibbons:1976ue}, where $\Kag_0$ is the extrinsic curvature of the flat background metric in which the full spacetime must be isometrically embedded \cite{Hawking:1995fd}. 

It is possible to reformulate general relativity in terms of tetrads\footnote{To be precise, $h^a{}_\mu$ are the components of co-tetrads, but it is common to call them tetrads as well.} $h^a{}_\mu$ and Riemannian spin connection $\ombol^a{}_{b \mu}$  instead of the metric $g_{\mu\nu}$ and Riemannian linear connection  $\Gammabol^\rho{}_{\nu\mu}$. The tetrads are related to the metric tensor as
\begin{equation}
	g_{\mu \nu}=\eta_{a b}h^a{}_\mu h^b{}_\nu,
\end{equation}
where  $\eta_{a b}=\text{diag}(-1, 1, 1, 1)$ is the tangent space Minkowski metric, and the Riemannian spin connection is given in terms of tetrads as
\begin{equation}
	\ombol^a{}_{b \mu}=\frac{1}{2} h^c{}_\mu \left( f_b{}^a{}_c+f_c{}^a{}_b-f^a{}_{b c} \right),
\end{equation}
where $ f^c{}_{a b}=h_a{}^\mu h_b{}^\nu (\partial_\nu h^c{}_\mu - \partial_\mu h^c{}_\nu)$ are the coefficients of anholonomy.

From the perspective of differential geometry, the Riemannian geometry can be understood to be a special case of a metric-affine geometry characterized by vanishing torsion and non-metricity. We can  consider a complementary approach, known as the (metric) teleparallel geometry,  defined by  conditions of vanishing non-metricity and curvature, which imply that  the teleparallel spin connection $\omega^a{}_{b\mu}$  has a pure-gauge form
\begin{equation}\label{teleconn}
	\omega^a{}_{b \mu}=\Lambda^a{}_c \partial_\mu (\Lambda^{-1})^c{}_b,
\end{equation}
where $\Lambda^a{}_b\in SO(1,3)$ in the Lorentzian case or $\Lambda^a{}_b\in SO(4)$ in the Euclidean case,
and generally has a non-vanishing torsion
\begin{equation}\label{torsion}
	T^a{}_{\mu \nu} = \partial_\mu h^a{}_\nu - \partial_\nu h^a{}_\mu + \omega^a{}_{b \mu} h^b{}_\nu - \omega^a{}_{b \nu} h^b{}_\mu.
\end{equation}
The teleparallel connection can be related to the Riemannian spin connection using the identity
\begin{equation}\label{telspincondec}
	{\omega^a}_{b \mu}=\ombol^a{}_{b \mu}+{K^a}_{b \mu},
\end{equation}
where we have defined the so-called contortion tensor
\begin{equation}
	K^a{}_{b \mu}=\frac{1}{2}({{T_\mu}^a}_b+{{T_b}^a}_\mu-{T^a}_{b \mu}).
\end{equation}

It is  rather straightforward to show that applying the identity \eqref{telspincondec} to the Riemannian curvature scalar, we obtain the geometric identity \cite{Aldrovandi:2013wha}
\begin{equation}\label{geo_identity}
-\Rbol=T+\frac{2}{h}\partial_\mu ( h\,T^\mu),
\end{equation}
where $T^\mu=T^{\nu\mu}{}_\nu$ is the vectorial torsion and 
\begin{equation}
	\label{tscalar}
	T = \frac{1}{2}S_{\mu}{}^{\rho\sigma} T^{\mu}{}_{\rho\sigma}=\frac{1}{4} T^\rho{}_{\mu \nu} T_\rho{}^{\mu \nu}+\frac{1}{2} T^\rho{}_{\mu \nu} T^{\nu \mu}{}_\rho - T^\rho{}_{\mu \rho} T^{\nu \mu}{}_\nu,
\end{equation}
is the torsion scalar, and $S_\mu{}^{\rho\sigma}$ is the superpotential defined as 
\begin{equation}
S_\mu{}^{\rho\sigma}=\frac{1}{2}
\left(
T_\mu{}^{\rho\sigma}+
T^{\sigma\rho}{}_\mu
-T^{\rho\sigma}{}_\mu
\right)
-\delta_\mu^\sigma T^{\rho}
+\delta_\mu^\rho T^{\sigma}.
\end{equation}
We can then consider the teleparallel gravity action
\begin{equation}\label{tegr_action}
\Sag_\text{TG}(h^a{}_\mu,\omega^a{}_{b\mu})=-\frac{1}{2\kappa}\int_\Mag h\, T,	
\end{equation}
and from \eqref{geo_identity} follows that it is dynamically equivalent to general relativity, meaning that the vacuum field equations derived by varying \eqref{tegr_action} with respect to $h^a{}_\mu$
\begin{equation}\label{teleeq}
h^{-1}\partial_\sigma (h S_{\mu}{}^{\rho\sigma})+\kappa  t_\mu{}^{\rho}=0,
\end{equation}
where 
\begin{equation}\label{pseudo}
t_\mu{}^\rho=\frac{1}{\kappa}\Gamma^\alpha{}_{\sigma\mu}S_{\alpha}{}^{\sigma\rho}
+\frac{1}{2\kappa}\delta_{\mu}^\rho T,
\end{equation}
is the energy-momentum pseudo-tensor, 
are the same as Einstein field equations, i.e.
\begin{equation}\label{tggreq}
h^{-1}\partial_\sigma (h S_{\mu}{}^{\rho\sigma})+\kappa  t_\mu{}^{\rho}
\equiv
\Rbol_\mu{}^{\rho}-\frac{1}{2}\delta_\mu{}^{\rho} \Rbol.
\end{equation}

This equivalence shows that the teleparallel field equations \eqref{teleeq} do not determine all 16 components of the tetrad but only 10 components of the metric tensor. Similarly, the teleparallel spin connection generated by the 6 components of $\Lambda^a{}_b$ is not determined by any field equations, as it can be shown to form only a total derivative term in the action \cite{Krssak:2015lba}
\begin{equation}\label{rel}
\Lag_\text{TG}(h^a{}_\mu,\omega^a{}_{b\mu})	=
\Lag_\text{TG}(h^a{}_\mu,0)
+
\frac{1}{\kappa}\partial_\mu(h\ \omega^{\mu}),
\end{equation}
where $\omega^\mu=\omega^a{}_{b\mu}h_a{}^\nu h^{b\mu}$.

This means that, as far as the field equations are concerned, teleparallel gravity is equivalent to general relativity \eqref{tggreq}, while the local Lorentz degrees of freedom $\Lambda^a{}_b$, related to both the tetrad and the spin connection, are not determined by the field equations. However, since we are interested in the value of the (Euclidean) gravitational action, the total derivative term in \eqref{rel} plays a crucial role, and we must determine how to evaluate it. The value of the gravitational action \eqref{rel} therefore depends on both the tetrad and the spin connection, and on how they are matched. We refer to the pair $\{h^a{}_\mu,\tilde{\omega}^a{}_{b\mu}\}$ as the \textit{proper frame}. Determining the proper frame therefore amounts to fixing the boundary term in \eqref{rel}, which is effectively equivalent to determining the counterterm in general relativity \cite{Krssak:2015rqa,Krssak:2024kva}. For this reason, we refer to this procedure as \emph{teleparallel renormalization}.

Due to the pure-gauge nature of the teleparallel spin connection \eqref{teleconn}, one can always construct the so-called \textit{proper tetrad} by a simultaneous local Lorentz transformation of the proper frame
\begin{equation}\label{key}
	\{h^a{}_\mu,\tilde{\omega}^a{}_{b\mu}\}
	\xrightarrow{\tilde{\Lambda}^a{}_b}\{\tilde{h}^a{}_\mu,0\}.
\end{equation}
which  is often referred to as the Weitzenb\"ock gauge \cite{Obukhov:2002tm}. The possibility to eliminate the teleparallel spin connection, allows one to formulate teleparallel gravity entirely in terms of the tetrad, resulting in the so-called pure-tetrad formulation \cite{Maluf:2013gaa}.  Athough this formulation is not manifestly covariant and involves conceptual issues, such as identifying the tensorial torsion with the non-tensorial coefficients of anholonomy \cite{Krssak:2024kva}, it is effectively equivalent to the covariant formulation. In practice, both formulations require determining the same local Lorentz transformation $\tilde{\Lambda}^a{}_b$  corresponding to the tetrad $h^a{}_\mu$: in the pure-tetrad formulation it is used to construct the proper tetrad $\tilde{h}^a{}_\mu=\tilde{\Lambda}^a{}_b h^b{}_\mu $, while in the covariant formulation it determines the corresponding pure-gauge spin connection $\tilde{\omega}^a{}_b=\tilde{\Lambda}^a{}_c \partial_\mu (\tilde{\Lambda}^{-1})^c{}_b$.

\section{Euclidean  Actions for Schwarzschild Black Hole}
Our goal in this paper is to understand various choices of the spin connection, how uniquely they are determined, and how well they agree with the result of general relativity in the Euclidean case. We demonstrate this on the example of the Euclidean Schwarzschild black hole
\begin{equation}\label{euclidschw}
	ds^2=f^2d\tau^2 + f^{-2} dr^2 +r^2 d\Omega^2, \quad f^2 \equiv 1-\frac{2M}{r},
\end{equation}
obtained from the formal analytical continuation of the Lorentzian time
\begin{equation}\label{key}
t\rightarrow -i \tau\,.	
\end{equation}
This prescription is particularly natural for the Schwarzschild solution in spherical coordinates, since $t$ is the static Killing time normalized at infinity. The corresponding real Euclidean section is regular provided that the imaginary time is periodically identified as $\tau\sim\tau+\beta$, where $\beta=8\pi M$ is interpreted as the inverse temperature with respect to this asymptotically normalized Killing time~\cite{Gibbons:1976ue,York:1986it}. Nevertheless, the Wick rotation should not be regarded as a generally covariant operation on an arbitrary Lorentzian manifold. In generic curved spacetimes, it turns out that it is better understood in terms of a complex deformation of the metric than as an analytic continuation of a coordinate~\cite{Visser:2017atf}.

At the level of the path integral, periodicity in imaginary time provides the connection with the canonical partition function, while the analytic continuation changes the oscillatory Lorentzian phase into a Euclidean statistical weight~\cite{Hawking:1979ig,Gibbons:1976ue}. More precisely, the Euclidean action is defined by
\begin{equation}
S_E=-iS_L\big|_{t=-i\tau},
\end{equation}
so that
\begin{equation}
e^{iS_L[g_L]}\longrightarrow e^{-S_E[g_E]}.
\end{equation}
At leading semiclassical order, the contribution of the Euclidean Schwarzschild saddle to the canonical partition function is therefore
\begin{equation}
Z(\beta)\big|_{\text{Schw}}\approx
 e^{-S_E[g_E^{\text{Schw}}]}.
\end{equation}
In what follows, we evaluate this Euclidean action using both the background-subtraction method of Gibbons and Hawking and the teleparallel
action.

\subsection{Background Subtraction}
Let us briefly summarize the standard calculation using the original  background subtraction method introduced by Gibbons and Hawking \cite{Gibbons:1976ue}. To evaluate the action \eqref{actiongr} we find the normal vector to the surface of constant radial distance as $n^\mu = f \delta^{\mu r}$. The  determinant of the induced metric on  this surface and the trace of its extrinsic curvature are given by 
\begin{equation}\label{nvecschw}
\sqrt{\gamma}= f\,r^2 \sin{\theta}, \qquad
	\Kag = \frac{2r-3M}{r^2 f} .
\end{equation}
The extrinsic curvature of the reference background metric is $K_0=2/r$.
For Euclidean time with a period $\beta=8\pi M$ and asymptotically flat spacetime, where we send the boundary to infinity $r\rightarrow \infty$, we find the well-known Gibbons-Hawking result
\begin{equation}\label{GHYext}
 \Sag^E_{\text{GR}}=\frac{1}{2}
 \int_{0}^{\beta}d\tau\,  \left[ 3M-2\left(1-\sqrt{f}\right)r\right]_{r\rightarrow\infty} = \frac{\beta M}{2},
\end{equation}
from which we can straightforwardly derive the black hole entropy area law.

\subsection{Teleparallel Euclidean Actions \label{sec32}}
We are interested in the calculation of the Euclidean action in teleparallel gravity, where several additional issues, compared to the situation in general relativity, need to be addressed. First we need to specify how to perform the Wick rotation in tetrad formalism. We can consider either the tangent space Wick rotation \cite{Samuel:2015oea}, where the tangent space metric is Wick rotated to the Euclidean one, i.e. $\eta_{\mu\nu}\rightarrow \delta_{\mu\nu}$, or Wick rotate the spacetime metric simultaneously with the tangent metric, and then find a tetrad corresponding to these Euclidean metrics. While for the diagonal metrics and tetrads both methods are equivalent, in the non-diagonal case, we find that only the latter method works.

We also need to specify how to integrate the Lagrangian density on  the Euclidean Schwarzschild manifold $\Eag$, which represents only the exterior of a black hole. One method is to take the volume integral over the whole  Euclidean manifold
\begin{equation}\label{actionorig}
	{S}^E_\text{TG}=\frac{1}{2\kappa}\int_{\Eag}^{} h T
	=\frac{1}{2\kappa}\int_0^\beta d\tau\int_0^\pi d\theta \int_0^{2\pi}d \phi \int_{2M}^{\infty}dr\,  h T,
\end{equation}
which we shall refer to as the \textit{bulk action}.

However, for vacuum spacetimes, we have another  method of evaluating the gravitational action based on the fact that  the vacuum torsion scalar can be written as a divergence of the vectorial torsion
\begin{equation}\label{vacrel}
h\,T= - 2\,\partial_\mu (hT^\mu),	
\end{equation}
or
\begin{equation}\label{vacrel2}
T=-2 \nablabol_\mu T^\mu.	
\end{equation}
Using Stokes' theorem, we can then transform the bulk action \eqref{actionorig} into the \textit{quasi-local action} 
\begin{equation}\label{actiontrick}
	\tilde{\Sag}_\text{TG}^E=-
	\frac{1}{\kappa}\oint_{\partial \Eag}\sqrt{\gamma}
	n_\mu T^\mu,	
\end{equation}
where same as in the case of the GHY term, $\partial \Eag$ is the boundary of the Euclidean manifold, taken to be a surface of constant radial coordinate, which is then sent to infinity.

Let us remark here that the relation \eqref{vacrel} can be obtained using the geometric identity \eqref{geo_identity} and the well-known result of general relativity that the Riemannian curvature scalar vanishes for vacuum spacetimes \cite{Oshita:2017nhn,Fiorini:2023axr,BeltranJimenez:2018vdo}.  However, if we wish to view teleparallel gravity as a fundamental theory we should not rely on switching back to general relativity for a such central identity. It turns out that we can  derive \eqref{vacrel} entirely within teleparallel gravity by noticing that  it is a trace of the vacuum field equation \eqref{teleeq}. Indeed, we have $\kappa t_\mu^\mu= T$ and $S_\rho{}^{\rho\mu}=2T^\mu$.

\subsubsection*{Proper Frame \label{secpropframe}}
In order to evaluate the teleparallel action, we need to match the teleparallel spin connection to the tetrad; otherwise, the teleparallel action is IR divergent.
One  of the methods to determine the spin connection--usually known as the proper frame  \cite{Lucas:2009nq,Krssak:2015rqa}--is based on the fact that the arbitrary teleparallel connection \eqref{teleconn} can be written as a Levi-Civita connection of the flat reference tetrad $\tref^a{}_\mu$
\begin{equation}\label{teleconnref}
\omega^a{}_{b\mu}=\ombol^a{}_{b\mu}(\tref^a{}_\mu).
\end{equation}
Calculating the teleparallel spin connection \eqref{teleconnref} is then equivalent to finding the reference tetrad $\tref^a{}_\mu$ corresponding to the ``full" tetrad $h^a{}_\mu$, which can be understood as analogous to finding the reference metric in the background subtraction method.

In the case of Schwarzschild solution, one of the choices of the tetrad is  the diagonal one
\begin{equation}\label{tetschw}
	h^a{}_{\mu}=\text{diag}
	\left(
	f,f^{-1},r,r\sin\theta
	\right),	
\end{equation}
for which the action with zero connection, $\Sag_\text{TG}(h^a{}_\mu,0)$ diverges. Therefore, we consider the reference tetrad
\begin{equation}\label{tetref}
	\tref^a{}_{\mu}=\text{diag}
	\left(
	1,1,r,r\sin\theta
	\right),
\end{equation}
obtained as a limit $r\rightarrow\infty$ of \eqref{tetschw}, and find the corresponding spin connection using \eqref{teleconnref} as
\begin{equation}\label{spinconnec}
	\tilde{\omega}^1{}_{2 \theta}=-1, \quad  \tilde{\omega}^1{}_{3 \phi}=-\sin\theta, \quad  \tilde{\omega}^2{}_{3 \phi}=-\cos\theta.
\end{equation}
We can evaluate the bulk action 
\begin{equation}\label{bulkactionproper}
	\Sag^E_\text{TG}(h^a{}_\mu,\tilde{\omega}^a{}_{b\mu})=\frac{1}{2\kappa}\int_{\Eag}^{} h T=
	\frac{2}{\kappa}\int_{\Eag}^{} \frac{M+r(f-1)}{rf}  \sin \theta=
	\left.\int d\tau\, r(f-1)\right|^{r\rightarrow\infty}_{r=2M}=
	\beta M,
\end{equation}
where we find the result to be twice the standard result of general relativity.

On the other hand, if we decide to follow the quasilocal approach \eqref{actiontrick}, we find the vectorial torsion to be
\begin{equation}\label{vectorprop}
	T^\mu(h^a{}_\mu,\tilde{\omega}^a{}_{b\mu})=\left(
	0,\frac{-3M+2rf(1-f)}{r^2},0,0\right),
\end{equation}
where only the leading term in  the asymptotic expansion of the $r$-component is relevant for the calculation of the Euclidean action
\begin{equation}\label{largervector}
T^r(h^a{}_\mu,\tilde{\omega}^a{}_{b\mu})=-\frac{M}{r^2} +\mathcal{O}\left(\frac{1}{r^3}\right),
\end{equation}
from which we obtain the quasilocal action as
\begin{equation}\label{qlactionproper}
	\tilde{\Sag}_\text{TG}(h^a{}_\mu,\tilde{\omega}^a{}_{b\mu})=
	\frac{\beta M}{2},
\end{equation}
which agrees  with the standard result of general relativity obtained using the background subtraction method \eqref{GHYext}.

\subsubsection*{Canonical Frame}
The alternative choice of a frame was proposed in \cite{BeltranJimenez:2019bnx}, where it was argued that the frame with a vanishing gravitational energy-momentum pseudotensor
\begin{equation}\label{canframe}
t^\mu{}_\nu=0,
\end{equation}
has a certain privileged position and was named the canonical frame. Although originally this construction was done in  symmetric teleparallel gravity, it is possible to repeat the same construction within teleparallel gravity \cite{Frob:2021dmv,Stano:2025eje}.
The metric satisfying the condition \eqref{canframe} in symmetric teleparallel gravity is the Kerr-Schild class of spacetimes in the Cartesian coordinate system
\begin{equation}\label{kerrschild}
	g_{\mu \nu} = \eta_{\mu\nu} + 2Fk_\mu k_\nu, 
\end{equation}
where $k_\mu=(1,x^i/r)$ is the null vector, $r=\sqrt{x^2+y^2+z^2}$, and  $F=M/r$ in the  Schwarzschild case. 

We can then  construct the tetrad using the Cartesian coordinate system
\begin{equation}\label{ksrealtetCart}
	\,^\text{CC} h^a{}_\mu=\delta^a{}_\mu + F k_\mu k^a,	
\end{equation}	
or  consider this metric in the spherical coordinate system  and  construct  the tetrad as   \cite{Stano:2025eje}
\begin{equation}\label{ksrealtet}
\,^\text{C} h^a{}_\mu=\tref^a{}_\mu + F k_\mu k^a,	
\end{equation}
where $\tref^a{}_\mu$ is given by \eqref{tetref}, and hence $\,^\text{C} h^a{}_\mu$  corresponds to a tetrad adapted to the Schwarzschild metric in Eddington-Finkelstein coordinates.

We can now observe that, while \eqref{ksrealtetCart} is a canonical frame with vanishing spin connection, i.e., it satisfies the condition \eqref{canframe}, the tetrad \eqref{ksrealtet} is not canonical and leads to an IR-divergent action. The reason is that the two tetrads are related by a coordinate transformation and a local Lorentz transformation. The coordinate transformation has no effect on the value of the action, whereas the local Lorentz transformation is responsible for the appearance of a non-trivial spin connection \cite{Krssak:2015lba}. The spin connection associated with \eqref{ksrealtet} can therefore be obtained either from the local Lorentz transformation relating it to \eqref{ksrealtetCart} using \eqref{teleconn}, or by applying the prescription discussed previously in the case of the proper frame. In the limit $r\rightarrow\infty$, the tetrad \eqref{ksrealtet} approaches the same reference tetrad \eqref{tetref} as the diagonal tetrad, and hence we  associate with \eqref{ksrealtet}  the same spin connection \eqref{spinconnec} as in the case of the diagonal tetrad.
The canonical frame will be then $\{\,^\text{C}h^a{}_\mu,\tilde{\omega}^a{}_{b\mu}\}$, and we  can verify that the canonical frame condition \eqref{canframe}   is satisfied.

Let us now consider the Wick rotation, which must be approached cautiously since the metric \eqref{kerrschild} and the tetrad $\eqref{ksrealtet}$ are not diagonal. It turns out that the  tangent space Wick rotation, i.e. using \eqref{ksrealtet} and changing only the tangent space metric $\eta_{\mu\nu}\rightarrow \delta_{\mu\nu}$, does not work as the resulting frame is not a solution of the field equations. Therefore, we have to Wick rotate not only the tangent metric, but also  the Kerr-Schild metric \eqref{kerrschild},  and then find a tetrad adopted to this complexified metric as\footnote{
While complex metrics and tetrads are somewhat unorthodox, they are nevertheless known to yield correct results \cite{Kontsevich:2021dmb,Witten:2021nzp,Visser:2021ucg}. From the viewpoint of the Kerr–Schild construction, the Wick rotation is even more problematic, since Kerr–Schild spacetimes rely on null vectors, which do not exist in the Euclidean case. Therefore, it is more appropriate to interpret this procedure as the Wick rotation of the Schwarzschild metric in Eddington–Finkelstein coordinates, followed by the construction of the corresponding tetrad \eqref{ImagKSframe}.}
\begin{equation}\label{ImagKSframe}
	\,^\text{EC}h^a{}_\mu = 
	\begin{pmatrix}
		1-\frac{M}{r} & -\frac{iM}{r} & 0 & 0\\
		-\frac{iM}{r} & 1+\frac{M}{r} & 0 & 0\\
		0 & 0 & r & 0\\
		0 & 0 & 0 & r \sin \theta
	\end{pmatrix}.
\end{equation}
We can verify that the resulting Euclidean canonical frame $\{\,^\text{EC}h^a{}_\mu,\tilde{\omega}^a{}_{b\mu}\}$ satisfies the condition \eqref{canframe}. 

Calculating the bulk gravitational action for the Euclidean canonical frame, we find that it is identically zero
\begin{equation}\label{bulkactioncan}
\Sag_\text{TG}(\,^\text{EC}h^a{}_\mu,\tilde{\omega}^a{}_{b\mu})=0,
\end{equation}
while the vectorial torsion is found to be  non-zero
\begin{equation}\label{vectorcan}
	T^\mu(\,^\text{EC}h^a{}_\mu,\tilde{\omega}^a{}_{b\mu})=\left(
	\frac{iM}{r^2},-\frac{M}{r^2},0,0\right),
\end{equation}
where only the $r$-component will be relevant if we are interested in the surface of constant $r$. 
The normal vector to a surface of constant $r$ and the determinant of the induced metric  are 
\begin{equation}\label{key}
n_\mu = f^{-1}\delta_{\mu r}, \qquad\sqrt{\gamma}= f\,r^2 \sin{\theta}.
\end{equation}
We can then use the quasi-local formula \eqref{actiontrick} to find the action
as
\begin{equation}\label{qlactioncan}
		\tilde{\Sag}_\text{TG}(\,^\text{EC}h^a{}_\mu,\tilde{\omega}^a{}_{b\mu})=\frac{\beta M}{2},
\end{equation}
which agrees with the previous quasi-local results using the proper frame \eqref{qlactionproper} and the background subtraction method \eqref{GHYext}.

\section{Lorentzian Actions on Wheeler-de Witt Patch}
While the Euclidean actions are of our primary interest here, it is interesting to mention similar calculations in the Lorentzian case as well. Particularly interesting is the calculation of the action growth,  i.e., the time derivative of the gravitational action,  evaluated on the Wheeler-de Witt (WdW) patch,  corresponding to the interior region of a black hole. Such a gravitational action growth plays an important role  in the  ``complexity=action" conjecture, where it was conjectured to correspond to the holographic complexity \cite{Brown:2015bva,Brown:2015lvg}. For us, it is an interesting scenario to test our methods of calculation of the gravitational action.

In general relativity, the gravitational action growth of the Schwarzschild solution on the WdW patch $\mathcal{W}$ is obtained by treating both $r=0$ and $r=2M$ as boundaries and evaluating the GHY term on them\footnote{We remark here that no counterterm was considered in the original proposal \cite{Brown:2015bva,Brown:2015lvg}, which can be justified by the fact that the Gibbons-Hawking counterterm takes the form $\sqrt{-\gamma} \mathcal{K}_0=2rf \sin\theta$ and hence vanishes at both boundaries.}
\begin{equation}\label{key}
	\left.\frac{d \Sag_\text{GR}}{d t}\right|_\mathcal{W}=\frac{1}{\kappa} \oint_{\partial\mathcal{W}}	\left.\sqrt{-\gamma} \mathcal{K}\right|_{r=0}^{r=2M}
	=\frac{1}{\kappa} \oint_{\partial\mathcal{W}}	\left.\sin\theta(2r-3M)\right|_{r=0}^{r=2M}	=2M.
\end{equation}

In the teleparallel case, it was shown recently that the proper frame leads to the same result using the bulk action integrated over the  WdW patch \cite{Krssak:2023nrw}\footnote{In \cite{Krssak:2023nrw}, this agreement was demonstrated also for asymptotically AdS and charged black holes, where the WdW patch corresponds to the region between the outer and inner horizons, which are both treated as boundaries. 
}
\begin{equation}\label{key}
\left.\frac{d {\Sag}_\text{TG}(h^a{}_\mu,\tilde{\omega}^a{}_{b\mu})}{d t}\right|_\mathcal{W}=	\left. r(1-f)\right|^{r=2M}_{r=0}=2M.	
\end{equation}
We obtain the same result using the growth of the quasilocal action for the proper frame 
\begin{eqnarray}
	\left.\frac{d\tilde{\Sag}_\text{TG}(h^a{}_\mu,\tilde{\omega}^a{}_{b\mu})}{d t}\right|_\mathcal{W}=\left.\frac{1}{\kappa} \oint_{\partial\mathcal{W}} \sin \theta \left[M-2rf(1-f)\right]\right|_{r=0}^{r=2M}
=
2M.
\end{eqnarray}

On the other hand, the canonical frame fails to provide the standard result using both  bulk or quasilocal actions. The bulk action growth is identically zero \eqref{bulkactioncan}, as well as the growth of the quasilocal action 
\begin{eqnarray}
\left.\frac{d\tilde{\Sag}_\text{TG}(\,^\text{EC}h^a{}_\mu,\tilde{\omega}^a{}_{b\mu})}{d t}\right|_\mathcal{W}
	=-\left.\frac{1}{\kappa} \oint_{\partial\mathcal{W}}  \sin \theta M\right|_{r=0}^{r=2M}	=0.
\end{eqnarray}

\section{Gravitational Action and Singularity \label{secSing}}
Although the results presented in the previous two sections may at first appear inconsistent or even contradictory, they in fact highlight the crucial role of the spacetime singularity. Once this singularity is properly taken into account, these results become fully consistent with each other.

Let us recall that the quasilocal action was obtained using the identity \eqref{vacrel} and  Stokes' theorem. For the Euclidean action on a region $\mathcal{U}$, this can be written as
\begin{equation}\label{stokes}
\frac{1}{2\kappa}\int_\mathcal{U}h\, T=
-\frac{1}{\kappa}\int_\mathcal{U}\partial_\mu (h\,T^\mu)
=-
\frac{1}{\kappa}\oint_{\partial\mathcal{U}}\sqrt{\gamma}\, n_\mu T^\mu,		
\end{equation}
and the analogous relation holds for the Lorentzian action. Here, it should be emphasized that the last equality in \eqref{stokes} is valid only if $T^\mu$ is regular everywhere on $\mathcal{U}$, which  is violated for the vectorial torsions \eqref{vectorprop} and \eqref{vectorcan}, both of which are singular at $r=0$.

It is illustrative to recall the familiar example of the total charge of a point particle in electrostatics. Starting from Gauss's law in differential form, $\mathbf{\nabla}\cdot\mathbf{E}=\rho/\epsilon_0$, the total charge can be written as
\begin{equation}\label{key}
Q=\int_V \rho\, d^3x=\int_V \mathbf{\nabla}.\mathbf{E}  \,d^3x=\oint_{\partial V} \mathbf{E}.\mathbf{a},
\end{equation}
where $\textbf{a}$ is the outward-pointing area element.
For a point charge with the electric field given by
\begin{equation}\label{key}
\mathbf{E}=\frac{q}{4\pi\epsilon_0}\frac{\mathbf{r}}{r^3},	
\end{equation}
we find the last integral to give us the total charge $Q=q$, while $\mathbf{\nabla}.\mathbf{E}=0$. This difference is caused by the fact that the electric field of a point particle is singular at $r=0$, and the total charge is located at this singular point.

There are multiple ways  to approach this situation, the most well-known being to introduce the Dirac delta function and taking the divergences  in the distributional sense. An alternative is to remove the singular point, leaving a punctured space with nontrivial topology and cohomology \cite{Nakahara:2003nw,Baez:1995sj}. However, we can avoid these simply by regarding the volume integral  of $\mathbf{\nabla}.\mathbf{E}$ as  the total charge everywhere where the $\mathbf{E}$ is regular, i.e. $r\neq0$, which is indeed zero. The surface integral calculates the total charge inside the volume bounded by $\partial V$, which also includes the point $r=0$. Therefore, we can  regard both integrals to be meaningful but represent different physical situations depending on whether we want to include the singular charge of a point particle in our result or not.

Our calculation of the gravitational action \eqref{stokes} can be viewed in an analogous way, where the trace of the field equations \eqref{vacrel} and vectorial torsion play the roles of  Gauss's law and electric field, respectively. The surface integral in \eqref{stokes} calculates the gravitational action as the flow of the vectorial torsion through $\partial\mathcal{U}$, while the volume integral  calculates the action as an integral over $\mathcal{U}$. In the case of the Euclidean manifold, the bulk action \eqref{actionorig} is a volume integral over the region above the horizon, while the result from the quasi-local action \eqref{actiontrick} also includes the contribution of the region under the horizon, including the singularity.

We can show this explicitly by calculating the quasilocal action as a flow of the vectorial torsion between two surfaces taken to be surfaces of constant radial coordinate: at $r_A$ and $r_B$. In the Euclidean case, we consider $r_A=2M$  and send  $r_B\rightarrow\infty$, and we can verify that the result agrees with the bulk action \cite{Stano:2025eje,Kuntz:2024qgs}. When we remove the inner boundary at the horizon and consider the flow of vectorial torsion only through the surface $r_B\rightarrow\infty$, we include the contribution from under the horizon as well. Since the Euclidean manifold $\Eag$ covers only the region outside of the horizon, we cannot distinguish between contributions from the singularity at $r=0$ and the remainder of the region under the horizon.

On the other hand, in the Lorentzian case, we can probe the region under the horizon  directly, and show that the bulk action agrees with the  quasilocal action calculated as a flow between two surfaces at $r_A$ and  $r_B$, even when one or both surfaces lie inside the horizon\footnote{In the Lorentzian case, the time is not periodic like in the Euclidean case. In order to make these integral meaningful we have to either consider the action growth or consider the action evaluated only between two hypersurfaces $t_i$ and $t_f$. We should also remark that when we evaluate the action growth not over the whole WDW patch, there will be imaginary contribution \cite{Krssak:2023nrw}.}. Since the WdW patch has two boundaries located at the inner and outer horizons, this provides an explanation why the bulk action can be used in calculations of the holographic complexity \cite{Krssak:2024kva}. However, if we remove the inner boundary  at $r_A$, we obtain a disagreement with the bulk action, demonstrating  that  $r=0$ is exactly the point where the Stokes' theorem ``cannot be applied", because at this point the vectorial torsion is divergent and not a regular vector field. Then one can attempt to solve this problem by including the Dirac delta function \cite{Stano:2025eje}, or by treating the bulk and quasilocal actions as two different calculations, where the contribution of the singularity is only in the latter. 

Here we should remark that we  would like our action to capture the effect of singularity, and hence we should consider the quasilocal action \eqref{actiontrick}  to be the physically relevant quantity. This conclusion is motivated not only by the analogy with electromagnetism but also by examining the structure of the gravitational action \eqref{actiongr}. As is well known, the GHY term is required to cancel boundary contributions arising from variations of the derivatives of the metric. Without this term, these derivatives would have to be fixed at the boundary in addition to the metric itself, rendering the variational principle ill posed \cite{Poisson:2009pwt}. However, these boundary contributions are obtained  only after using the Palatini identity and converting a total divergence term into a surface term using Stokes' theorem. We usually consider only the outer boundary, and hence the action \eqref{actiongr} should be viewed as  incorporating  contributions associated with the singularity  enclosed by the boundary.

\section{Proper Frame Revisited\label{secpfrev}}
Let us now discuss the physical interpretation of the proper frame discussed in Section~\ref{secpropframe}, in which the teleparallel spin connection associated with a given tetrad is constructed as the Levi-Civita connection of the reference tetrad \eqref{teleconnref}. The original motivation for this prescription \cite{Krssak:2015rqa} was to ensure that the torsion represents only the gravitational field and therefore vanishes in the absence of gravity. For asymptotically flat spacetimes, this implies that the torsion should vanish at spatial infinity. However, as we will show, this requirement is not sufficient: the torsion must not only vanish asymptotically but do so sufficiently rapidly.

As we have seen, this prescription is required for both the diagonal tetrad \eqref{tetschw} and the Eddington--Finkelstein tetrad \eqref{ksrealtet} (or \eqref{ImagKSframe} in the Euclidean case). In both cases, it yields the same spin connection \eqref{spinconnec}, reflecting the fact that the two tetrads approach the same reference tetrad asymptotically. This is sufficient because the finiteness of the action depends only on the asymptotic behavior of the torsion. Indeed, since $h=r^2\sin\theta=n_r\sqrt{|\gamma|}$, finiteness of the bulk action requires the torsion scalar to decay more rapidly than  $\mathcal{O}(r^{-3})$,  
while the quasilocal action is finite and non-vanishing when the radial component of the vectorial torsion behaves as $T^r=\mathcal{O}(r^{-2})$. A vectorial torsion with a faster fall-off, such as $T^r=\mathcal{O}(r^{-3})$, leads to a vanishing contribution at spatial infinity and, consequently, a zero quasi-local action.  

We can observe that, for the diagonal tetrad \eqref{tetschw} without the spin connection, the torsion falls off at large $r$ as 
$\mathcal{O}\left(1\right)$, as can be seen from the asymptotic behaviour of its non-vanishing components 
\begin{equation}\label{asymbad}
    T^t{}_{tr}\sim -\frac{M}{r^2}+\mathcal{O}(r^{-3}), \quad T^\theta{}_{r\theta}=T^\phi{}_{r\phi}=\frac{1}{r}, \quad T^\phi{}_{\theta \phi}=\cot{\theta},
\end{equation}
leading to the torsion scalar and radial component of  vectorial torsion fall-off behaviour as
\begin{equation}\label{}
T(h^a{}_{\mu},0)=-\frac{2}{r^2},
\qquad
T^r(h^a{}_{\mu},0)=\frac{2}{r} +	\mathcal{O}\left(\frac{1}{r^2}\right),	
\end{equation}
which obviously leads to  both the bulk and quasilocal actions being divergent.

We can then verify that  using the appropriate spin connection with the tetrad improves the fall-off behaviour to
\begin{equation}\label{key}
	T(h^a{}_{\mu},\omega^a{}_{b\mu})=\frac{2M^2}{r^4}+\mathcal{O}\left(\frac{1}{r^5}\right),
	\qquad
	T^r(h^a{}_{\mu},\omega^a{}_{b\mu})=
	-\frac{M}{r^2}+	\mathcal{O}\left(\frac{1}{r^3}\right),	
\end{equation}
which then lead to finite bulk and quasilocal actions.

Moreover, it turns out that the spin connection improves the overall asymptotic behaviour of the the torsion to $\mathcal{O}\left(r^{-2}\right)$, since  the asymptotic behaviour of the non-vanishing components of torsion is now given by
\begin{equation}
    T^t{}_{tr}=T^\theta{}_{r \theta}=T^\phi{}_{r\phi}\sim -\frac{M}{r^2}+\mathcal{O}(r^{-3}).
\end{equation}
Let us make several observations that clarify why this change in the asymptotic behavior of the torsion is crucial to understanding teleparallel renormalization. First, it is the asymptotic fall-off of the torsion tensor itself that is relevant. The source of the problematic behavior in \eqref{asymbad} can be traced to the tetrad covectors $h^{\hat{2}}=rd\theta$ and $h^{\hat{3}}=r\sin\theta d\phi$, which grow linearly with $r$ and consequently give rise to $\mathcal{O}\left(1\right)$ terms in the torsion. Crucially, this behavior is a consequence of the choice of tetrad and can be removed by introducing a teleparallel spin connection that accounts for the corresponding inertial effects \cite{Krssak:2015rqa,Krssak:2024kva}. This explains why the same spin connection is used for the diagonal tetrad \eqref{tetschw}, the Kerr--Schild tetrads \eqref{ksrealtet} and \eqref{ImagKSframe}, the freely falling tetrads \cite{Emtsova:2021ehh}, and the diagonal tetrad in Minkowski spacetime: all these tetrads contain the same $h^2$ and $h^3$, and hence the same spin connection \eqref{spinconnec} removes the problematic asymptotic behavior originating from them. The alternative is to consider tetrads that lead to a well-behaved torsion tensor without the spin connection, such as the Eddington--Finkelstein tetrad in Cartesian coordinates \eqref{ksrealtetCart} or the  proper tetrad introduced in \cite{Maluf:1995re}.
\section{Uniqueness of Teleparallel Renormalization \label{secuniq}}
Having determined the proper frame as a pair $\{h^a,\omega^a{}_b\}$,  leading  to a finite gravitational action, it is interesting to ask how unique this matching of the spin connection to the tetrad is. Since the action is invariant under simultaneous transformation of both the tetrad and spin connection, we  need to consider a fixed  tetrad and vary only the spin connection. Transformations of the spin connection  leaving the action unchanged represent the redundancy up to which the spin connection is determined for the proper frame.

Let us consider the $r$-dependent $SO(4)$ rotations 
\begin{equation}\label{Lamdaboost}
	\bar{\Lambda}^a {}_b =
	\left(
	\begin{array}{cccc}
		\cos \alpha & \sin \alpha & 0 & 0 \\
		-\sin\alpha  & \cos\alpha & 0 & 0 \\
		0 & 0 & 1 & 0 \\
		0 & 0 & 0 & 1 \\
	\end{array}
	\right),
\end{equation}
which can be considered to be an Euclidean version of a Lorentzian boost with rapidity $\alpha$. 

We then transform the spin connection $\omega\rightarrow \omega'$ while keeping the tetrad unchanged. Since the quasilocal action is given by the asymptotic value of the vectorial torsion, we need to examine the asymptotic behaviour of the rapidity $\alpha$. Writing down the leading term in the asymptotic expansion of the rapidity as
\begin{equation}\label{rapexp}
\alpha \sim \frac{\alpha_0}{r^p}, \qquad r\rightarrow \infty,
\end{equation}
we can then classify the radial Euclidean boosts into three categories: 
\begin{enumerate}
	\item $p>1/2$: leave the action invariant.
	\item $p<1/2$: cause the action to diverge.
	\item $p=1/2$: cause a change in the leading term in the vectorial torsion 
	\begin{equation}\label{key}
		T^r(h^a,\omega'^a{}_b)=\frac{\alpha_0^2-M}{r^2}+	\mathcal{O}\left(\frac{1}{r^3}\right),	
	\end{equation}
which  shifts the value of the quasilocal action   to 
\begin{equation}\label{qlactionpropershift}
	\tilde{\Sag}_\text{TG}(h^a{}_\mu,\tilde{\omega}'^a{}_{b\mu})=
	\frac{\beta}{2}\left(
    M-\alpha_0^2 \right).
\end{equation}
\end{enumerate}

It is straightforward to see that  transformations of the first kind are genuine symmetries of the gravitational action, while transformations of the second kind are not admissible. The central question is whether the third category of transformations should be regarded as symmetries of gravitational action. 

We would like to emphasize that transformations of the third kind are important as they include the Euclidean version of the boost that transforms between the diagonal and free-falling frames. In the Lorentzian case,  this transformation is given by a boost with rapidity $\alpha=\text{arccosh}(1/f)$. The Euclidean analogue of a free-falling frame is obtained  by using the $SO(4)$ rotation \eqref{Lamdaboost} with the rapidity $\alpha=\arccos(1/f)$. Since asymptotically  $\arccos(1/f)\approx \sqrt{-2M/r} +\mathcal{O}(r^{3/2})$, this results in the Euclidean action to be $3\beta M/2$, i.e., three times the expected value\footnote{The same result was obtained in \cite{Fiorini:2023axr}, where the authors considered a proper tetrad corresponding to the freely falling frame rather than using the spin connection. However, a closer inspection reveals that the Lorentzian tetrad was used  in the Euclidean calculation, see (3.12) in \cite{Fiorini:2023axr}. As we have discussed in the case of the canonical frame in Section~\ref{sec32}, this is is potentially problematic, and hence we argue in favor of  applying the $SO(4)$  rotation \eqref{Lamdaboost} rather than using the Lorentzian boosted-tetrad.}. 

From \eqref{qlactionpropershift} follows that Euclidean boosts with $p=1/2$ shift the value of the action by $\alpha_0^2$, and hence the Euclidean action can take practically an arbitrary value. This causes a problem with the uniqueness of the black hole thermodynamics, and hence it was suggested that such transformations should be excluded \cite{Fiorini:2023axr}. 

We would like to provide another argument for excluding these transformations, based on the observation made at the end of Section~\ref{secpfrev} that teleparallel renormalization improves the asymptotic behaviour of the torsion. We find that only Euclidean boosts \eqref{Lamdaboost} with rapidity \eqref{rapexp} corresponding to $p=1$, or, more generally, local Lorentz transformations that asymptotically behave as
\begin{equation}\label{LambdaAsym}
\Lambda^a{}_b=\delta^a_b +\mathcal{O}\left(\frac{1}{r}\right),
\end{equation}
preserve the asymptotic $\mathcal{O}(r^{-2})$ behavior of the torsion tensor. Thus, if we require the torsion tensor to fall off as $\mathcal{O}(r^{-2})$, only transformations of the form \eqref{LambdaAsym} are admissible, leading to a unique value of the Euclidean gravitational action.

This raises the question of why the torsion should be required to fall off asymptotically as $\mathcal{O}(r^{-2})$, and such a requirement can be justified. One argument, closely related to the original idea of \cite{Krssak:2015rqa},  is that the spin connection represents inertial effects that are present in the theory due to the equivalence principle. Although these inertial effects are locally indistinguishable from gravitational ones, they are distinguishable globally, i.e., by their behavior at spatial infinity. Inertial effects associated with the choice of frame need not vanish asymptotically, whereas the gravitational field as a physical field should decay sufficiently far from its source. For physical fields such as electromagnetism, the $\mathcal{O}(r^{-2})$ fall-off is required to recover the Coulomb solution and ensure finite conserved charges. Similarly, for a spherically symmetric gravitational field, recovering the Newtonian limit and finite conserved charges requires the torsion tensor to exhibit the $\mathcal{O}(r^{-2})$ asymptotic fall-off.

An important property of transformations \eqref{LambdaAsym} is that the product of any two such transformations has the same asymptotic fall-off. Thus, transformations of the form \eqref{LambdaAsym} are closed under composition and form a group, 
as expected for physical symmetries. This is in contrast to the so-called remnant symmetries discovered in the modified $f(T)$ gravity case \cite{Ferraro:2014owa,Ong:2013qja,Chen:2014qtl}, which are those local Lorentz transformations that leave the torsion scalar invariant. The problem of these remnant symmetries is that they do not form a group, i.e., combining two different symmetries is not a symmetry, which raises questions about their interpretation as physical symmetries \cite{Chen:2014qtl,Golovnev:2020zpv,Golovnev:2020nln}. 

Therefore, the transformations \eqref{LambdaAsym} are symmetries of teleparallel renormalization, representing a non-uniqueness in determining the proper frame $\{h^a{}_\mu,\tilde{\omega}^a{}_{b\mu}\}$, i.e., a residual freedom to perform local Lorentz transformations of the spin connection associated with a given tetrad. In the non-covariant formulation, these are the allowed transformations of the  proper tetrad $\tilde{h}^a{}_\mu$ that leave the quasilocal action invariant and are the example of asymptotically global Lorentz transformations introduced in \cite{Obukhov:2006sk}.  It is interesting to mention that the metric analogue of these transformations are the symmetries of the Einstein action discussed already by Fadeev \cite{Fadeev}, and are related to the asymptotic symmetries discussed in the Hamiltonian formulation   \cite{Regge:1974zd}.

\section{Discussion and Conclusions}
In this paper, we have revisited the problem of renormalizing the gravitational action in the teleparallel equivalent of general relativity. Although the teleparallel action does not require GHY boundary terms, it depends on the choice of the pure-gauge spin connection that enters the action only through the boundary term. Determining the proper frame, i.e., finding the spin connection associated with a given tetrad, fixes the boundary term and is therefore effectively equivalent to determining the counterterm in the standard formulation of general relativity. We have referred to this procedure as the teleparallel renormalization of the gravitational action and investigated it in detail for the Schwarzschild solution.

The main goal of this work was to clarify several apparently contradictory results concerning the value of the teleparallel gravitational action. We have shown that the main source of discrepancies is due to two different methods of evaluating the teleparallel action: the bulk action is obtained directly by integrating the Lagrangian over spacetime, whereas the quasilocal action follows after rewriting it as a surface term using the trace of the vacuum field equations and Stokes' theorem. Although these two actions are equivalent when torsion is regular over the whole spacetime, this turns out to be not the case in physically interesting cases.  

A useful analogy that sheds light on this issue is provided by Gauss's law for the electric field of a point charge: the flux through a closed surface enclosing the charge is nonzero, even though the divergence of the electric field vanishes everywhere in the regular region. As is well known, this can be resolved either by introducing a Dirac delta source or, equivalently, by observing that removing the singular point leaves a punctured space with nontrivial topology and cohomology. The difference between the volume and surface integrals can then be viewed as a method of detecting this nontrivial topology, i.e., the ``holes'' in our space.

For black-hole spacetimes, the situation is more involved, since there are two possible ``holes" that may play a role. In the Euclidean case, after imposing periodicity in Euclidean time, we are left with the Euclidean manifold $\Eag$, which corresponds only to the exterior region of the black hole. The removed black-hole interior can be viewed as a ``hole" in the topological sense, since $\Eag$ is not contractible to a point and is responsible for the discrepancy between the bulk and quasilocal actions. We have argued that these two actions can be understood as describing two different physical situations: the bulk action corresponds to the exterior region of the black hole only, whereas the quasilocal action represents the action of the entire region bounded by the surface at infinity and therefore includes the contribution from the black-hole interior as well. We have argued in favor of the quasilocal action, since in standard general relativity the action is determined by boundary terms evaluated only on the outer boundary at infinity.

In the Lorentzian case, the interior of the black hole is part of the Lorentzian manifold, allowing the under-horizon geometry to be probed directly. The only ``hole'' in the Lorentzian manifold is then the central singularity at $r=0$, where the Schwarzschild solution breaks down. We have indeed observed that the central singularity is the only source of the discrepancy between bulk and quasilocal actions in the Lorentzian case. When the central singularity is excluded, as is the case of  calculating  the action growth on the WdW patch in the CA duality proposal, the bulk and quasilocal actions agree. This finally provided a satisfactory explanation for our recent result \cite{Krssak:2018ywd}, in which it was  shown that the teleparallel bulk action reproduces the standard general-relativistic results.

The secondary goal of this paper was to understand the dependence of teleparallel renormalization on the proper frame construction. This was motivated by recent claims that the so-called canonical frame--given by the condition of vanishing gravitational energy-momentum pseudo-tensor \eqref{canframe}--leads uniquely to correct results for the energy-momentum and Euclidean actions 
\cite{BeltranJimenez:2024ufa}.  Let us comment on these claims from the perspective of the results of this paper.

Firstly, making the existence of the canonical frame a central ingredient of the theory raises the question of whether such a frame can always be realized. In the ordinary teleparallel gravity considered here, we have 6 local Lorentz degrees of freedom in choosing the tetrad, while the  canonical frame is given by 16 conditions \eqref{canframe}, and hence the canonical frame cannot always be realized. This led the authors of this proposal to generalize teleparallel theories and replace $SO(1,3)$ (or $SO(4)$) by a general $GL(4)$ pure-gauge connection \cite{Adak:2023ymc,Gomes:2023hyk}. While this is an interesting idea in general, the only known example that uses this $GL(4)$ connection is the recently proposed cosmological one \cite{Gomes:2023hyk}. All the other examples of canonical frame--which include all black holes solutions--reduce to the Kerr-Schild ansatz in either the ordinary or  symmetric teleparallel framework\footnote{Another interesting example 
that satisfies \eqref{canframe} is the recently found self-excited instanton solution \cite{Krssak:2024vzo}.}. 
This makes it rather unclear which of the attractive  properties of the canonical frame are specific to the Kerr–Schild ansatz, and one should exercise caution when promoting them to be properties of the theory itself.
\footnote{Note that the  energy-momentum pseudotensor vanishes and Einstein field equations linearize in the Kerr-Schild case was observed already in 1970s \cite{Gurses:1975vu}, and was used to calculate conserved quantities  in standard general relativity \cite{Virbhadra:1990zr}.}

Secondly, the idea of canonical frame was motivated by the equivalence principle and the existence of the inertial frame \cite{Koivisto:2022nar}. However, the equivalence principle--at least as understood in general relativity--requires the existence of only a locally inertial frame, while the canonical frame defined by \eqref{canframe}  concerns the existence of a global inertial frame.
While we do not see a compelling reason why the globally inertial frame should be  an essential ingredient of the theory itself, it is nevertheless interesting to investigate specific solutions that admit such a frame, most notably all solutions that can be written in the Kerr--Schild form. These solutions exhibit a number of remarkable properties, including  being an intriguing analogue of the point charge in electromagnetism, since the nontrivial contribution to the Euclidean action originates solely from the central singularity.
From a mathematical perspective, this is particularly interesting, as the vectorial torsion of the canonical frame is a closed but not globally exact form, allowing the interplay between the bulk and quasilocal actions to be understood in terms of cohomology.

Lastly, while the canonical frame agrees with the proper diagonal frame and the standard general relativity calculations for the Euclidean actions and energy-momentum, it does fail on the WdW patch, where the Lorentzian action growth vanishes identically. In contrast, the proper diagonal frame fully agrees with the standard general relativity results not only in the Schwarzschild case, but also in the charged and AdS cases as well \cite{Krssak:2023nrw}. While the CA duality is a conjecture, it nevertheless provides an interesting way to test the methods of evaluating the gravitational action.  The failure of the canonical frame to reproduce these results therefore demonstrates that it does not always lead to the correct results.

The final goal of this paper was to understand the uniqueness of teleparallel renormalization, i.e., whether a specific spin connection is uniquely associated with a given tetrad or whether there is a freedom to associate different spin connections with the same tetrad.  We have shown the latter to be the case and found freedom to perform residual local Lorentz transformations, which can be classified according to their asymptotic behaviour. Transformations that decay sufficiently rapidly--found to be boosts with rapidity decaying asymptotically faster than $1/\sqrt{r}$--leave the action unchanged and represent genuine symmetries of the renormalization procedure. 

We were then led to the question of whether boosts with rapidity falling off precisely as $1/\sqrt{r}$  should  be  regarded as symmetries of the renormalization procedure.
If such boosts are allowed, they shift the value of the action by a constant and consequently raise the possibility that the Euclidean action could be assigned an arbitrary value.  We have argued that if we require not only the vectorial torsion but all components of the torsion tensor to have asymptotic behaviour $\mathcal{O}(r^{-2})$, such boosts are naturally excluded and we obtain unique predictions for the value of the Euclidean action. It turns out that these symmetries of the teleparallel renormalization scheme are closely related to the problem of remnant symmetries in teleparallel theories, as well as to the symmetries of the Einstein action discussed by Faddeev \cite{Fadeev} and the asymptotic symmetries in the Hamiltonian formulation \cite{Regge:1974zd}.

Our results establish teleparallel renormalization as a consistent alternative to the standard methods of renormalizing the gravitational action. While we have focused here on the Schwarzschild solution, we have clarified the roles of singularities, the choice of proper frame, and residual local Lorentz freedom in determining the resulting action. This provides a basis for extending the analysis to more general spacetimes and for further investigating the relation between teleparallel renormalization, conserved charges, and the global structure of spacetime.

\section{Acknowledgements}
The work of MK was supported by  VEGA
grant 1/0565/25. MS has been supported by the Institute of Mathematics of the Czech Academy of Sciences (RVO 67985840) and the grant  GAČR 25-15544S.

\bibliography{references}

@article{Krssak:2023nrw,
	author = "Kr\v{s}\v{s}\'ak, Martin",
	title = "{Bulk action growth for holographic complexity}",
	eprint = "2308.04354",
	archivePrefix = "arXiv",
	primaryClass = "hep-th",
	doi = "10.1103/PhysRevD.109.086002",
	journal = "Phys. Rev. D",
	volume = "109",
	number = "8",
	pages = "086002",
	year = "2024"
}

@article{LeDelliou:2019esi,
    author = "Le Delliou, M. and Huguet, E. and Fontanini, M.",
    title = "{Teleparallel theory as a gauge theory of translations: Remarks and issues}",
    eprint = "1910.08471",
    archivePrefix = "arXiv",
    primaryClass = "gr-qc",
    doi = "10.1103/PhysRevD.101.024059",
    journal = "Phys. Rev. D",
    volume = "101",
    number = "2",
    pages = "024059",
    year = "2020"
}

@article{Fontanini:2018krt,
    author = "Fontanini, M. and Huguet, E. and Le Delliou, M.",
    title = "{Teleparallel gravity equivalent of general relativity as a gauge theory: Translation or Cartan connection?}",
    eprint = "1811.03810",
    archivePrefix = "arXiv",
    primaryClass = "gr-qc",
    doi = "10.1103/PhysRevD.99.064006",
    journal = "Phys. Rev. D",
    volume = "99",
    number = "6",
    pages = "064006",
    year = "2019"
}

@article{Pereira:2019woq,
    author = "Pereira, Jos\'e G. and Obukhov, Yuri N.",
    title = "{Gauge Structure of Teleparallel Gravity}",
    eprint = "1906.06287",
    archivePrefix = "arXiv",
    primaryClass = "gr-qc",
    doi = "10.3390/universe5060139",
    journal = "Universe",
    volume = "5",
    number = "6",
    pages = "139",
    year = "2019"
}

@article{Krssak:2024vzo,
    author = "Kr{\v{s}}{\v{s}}{\'a}k, Martin",
    title = "{Self-Excited Gravitational Instantons}",
    eprint = "2408.01140",
    archivePrefix = "arXiv",
    primaryClass = "gr-qc",
    doi = "10.1103/PhysRevLett.134.201501",
    journal = "Phys. Rev. Lett.",
    volume = "134",
    number = "20",
    pages = "201501",
    year = "2025"
}

@article{Kuntz:2024qgs,
    author = "Kuntz, Iber\^e and Paci, Gregorio and Zanusso, Omar",
    title = "{Euclidean actions and static black hole entropy in teleparallel theories}",
    eprint = "2410.04510",
    archivePrefix = "arXiv",
    primaryClass = "gr-qc",
    doi = "10.1088/1361-6382/ada867",
    journal = "Class. Quant. Grav.",
    volume = "42",
    number = "4",
    pages = "045011",
    year = "2025"
}

@article{Krssak:2018ywd,
    author = {Kr\v{s}\v{s}\'ak, M. and van den Hoogen, R.J. and Pereira, J.G. and B\"ohmer, C.G. and Coley, A.A.},
    title = "{Teleparallel theories of gravity: illuminating a fully invariant approach}",
    eprint = "1810.12932",
    archivePrefix = "arXiv",
    primaryClass = "gr-qc",
    doi = "10.1088/1361-6382/ab2e1f",
    journal = "Class. Quant. Grav.",
    volume = "36",
    number = "18",
    pages = "183001",
    year = "2019"
}

@article{Fiorini:2023axr,
    author = "Fiorini, Franco and Gonz\'alez, P. A. and V\'asquez, Yerko",
    title = "{Reference frames and black hole thermodynamics}",
    eprint = "2309.06293",
    archivePrefix = "arXiv",
    primaryClass = "gr-qc",
    doi = "10.1088/1475-7516/2023/12/033",
    journal = "JCAP",
    volume = "12",
    pages = "033",
    year = "2023"
}

@article{Krssak:2024kva,
    author = "Kr\v{s}\v{s}\'ak, Martin",
    title = "{Einstein gravity from the Einstein action: Counterterms and covariance}",
    eprint = "2406.08452",
    archivePrefix = "arXiv",
    primaryClass = "gr-qc",
    doi = "10.1103/PhysRevD.110.104061",
    journal = "Phys. Rev. D",
    volume = "110",
    number = "10",
    pages = "104061",
    year = "2024"
}

@article{Golovnev:2023yla,
    author = "Golovnev, Alexey",
    title = {{The geometrical meaning of the Weitzenb{\"o}ck connection}},
    eprint = "2302.13599",
    archivePrefix = "arXiv",
    primaryClass = "gr-qc",
    doi = "10.1142/S0219887823502195",
    journal = "Int. J. Geom. Meth. Mod. Phys.",
    volume = "20",
    number = "Supp01",
    pages = "2350219",
    year = "2023"
}

@article{Virbhadra:1990zr,
	author = "Virbhadra, K. S.",
	title = "{Energy distribution in Kerr-Newman space-time in Einstein's as well as Moller's prescriptions}",
	doi = "10.1103/PhysRevD.42.2919",
	journal = "Phys. Rev. D",
	volume = "42",
	pages = "2919--2921",
	year = "1990"
}

@article{Golovnev:2017dox,
    author = "Golovnev, Alexey and Koivisto, Tomi and Sandstad, Marit",
    title = "{On the covariance of teleparallel gravity theories}",
    eprint = "1701.06271",
    archivePrefix = "arXiv",
    primaryClass = "gr-qc",
    reportNumber = "NORDITA-2017-11",
    doi = "10.1088/1361-6382/aa7830",
    journal = "Class. Quant. Grav.",
    volume = "34",
    number = "14",
    pages = "145013",
    year = "2017"
}

@article{Moller1966,
	author         = "M{\o}ller, Cristian",
	title          = "{Survey of Investigations on the Energy-Momentum Complex in General Relativity}",
	journal        = "K. Dan. Vidensk. Selsk. Mat. Fys. Skr. ",
	volume         = "35",
	year           = "1966",
	number         = "3",
	pages          = "1-14",
	
}

@article{Moller1978,
	author         = "M{\o}ller, Cristian",
	title          = "{On the Crisis in the Theory of Gravitation and a Possible Solution}",
	journal        = "K. Dan. Vidensk. Selsk. Mat. Fys. Skr. ",
	volume         = "89",
	year           = "1978",
	number         = "13",
	pages          = "1-32",
	
}

@article{Pellegrini1963,
	author         = "Pellegrini, C. and Plebanski, J. ",
	title          = "{Tetrad Fields and Gravitational Fields}",
	journal        = "K. Dan. Vidensk. Selsk. Mat. Fys. Skr. ",
	volume         = "2",
	year           = "1963",
	number         = "4",
	pages          = "1-39",
	
}

@book{Fecko:2006zy,
    author = "Fecko, M.",
    title = "{Differential geometry and Lie groups for physicists}",
    isbn = "978-0-521-18796-1, 978-0-521-84507-6, 978-0-511-24296-0",
    publisher = "Cambridge University Press",
    month = "",
    year = "2011"
}

@book{Baez:1995sj,
    author = "Baez, J. and Muniain, J. P.",
    title = "{Gauge fields, knots and gravity}",
    isbn = "978-7-301-22787-9",
    publisher = "World Scientific",
    year = "1995"
}

@article{Gurses:1975vu,
    author = "Gurses, Metin and Gursey, Feza",
    title = "{Lorentz Covariant Treatment of the Kerr-Schild Metric}",
    doi = "10.1063/1.522480",
    journal = "J. Math. Phys.",
    volume = "16",
    pages = "2385",
    year = "1975"
}

@article{Oshita:2017nhn,
	author = "Oshita, Naritaka and Wu, Yi-Peng",
	title = "{Role of spacetime boundaries in Einstein's other gravity}",
	eprint = "1705.10436",
	archivePrefix = "arXiv",
	primaryClass = "gr-qc",
	doi = "10.1103/PhysRevD.96.044042",
	journal = "Phys. Rev. D",
	volume = "96",
	number = "4",
	pages = "044042",
	year = "2017"
}

@book{Poisson:2009pwt,
    author = "Poisson, Eric",
    title = "{A Relativist's Toolkit: The Mathematics of Black-Hole Mechanics}",
    doi = "10.1017/CBO9780511606601",
    publisher = "Cambridge University Press",
    month = "",
    year = "2009"
}

@article{BeltranJimenez:2024ufa,
    author = "Beltr{\'a}n Jim{\'e}nez, Jose and Koivisto, Tomi S.",
    title = "{Euclidean teleparallel relativity and black hole partition functions}",
    eprint = "2412.13946",
    archivePrefix = "arXiv",
    primaryClass = "gr-qc",
    doi = "10.1103/5wtj-7k2c",
    journal = "Phys. Rev. D",
    volume = "113",
    number = "2",
    pages = "L021502",
    year = "2026"
}

@incollection{Krssak:2024xeh,
    author = "Kr\v{s}\v{s}\'ak, Martin",
    title = "{Teleparallel Gravity, Covariance and Their Geometrical Meaning}",
    booktitle = "{Tribute to Ruben Aldrovandi}",
    editor="{F. Caruso, J.G. Pereira and A. Santoro}",
    publisher = "{Editora Livraria da Física}",
    eprint = "2401.08106",
    archivePrefix = "arXiv",
    primaryClass = "gr-qc",
    address="São Paulo",
    month = "",
    year = "2024"
}

@article{Hawking:1995fd,
    author = "Hawking, S. W. and Horowitz, Gary T.",
    title = "{The Gravitational Hamiltonian, action, entropy and surface terms}",
    eprint = "gr-qc/9501014",
    archivePrefix = "arXiv",
    reportNumber = "DAMTP-R-94-52, UCSBTH-94-37",
    doi = "10.1088/0264-9381/13/6/017",
    journal = "Class. Quant. Grav.",
    volume = "13",
    pages = "1487--1498",
    year = "1996"
}

@article{York:1972sj,
    author = "York, Jr., James W.",
    title = "{Role of conformal three geometry in the dynamics of gravitation}",
    doi = "10.1103/PhysRevLett.28.1082",
    journal = "Phys. Rev. Lett.",
    volume = "28",
    pages = "1082--1085",
    year = "1972"
}

@article{York:1986it,
    author = "York, Jr., James W.",
    title = "{Black hole thermodynamics and the Euclidean Einstein action}",
    doi = "10.1103/PhysRevD.33.2092",
    journal = "Phys. Rev. D",
    volume = "33",
    pages = "2092--2099",
    year = "1986"
}

@article{Koivisto:2022nar,
    author = "Koivisto, Tomi",
    title = "{Energy in the relativistic theory of gravity}",
    eprint = "2202.11522",
    archivePrefix = "arXiv",
    primaryClass = "gr-qc",
    doi = "10.1142/S0219887822400072",
    journal = "Int. J. Geom. Meth. Mod. Phys.",
    volume = "19",
    number = "Supp01",
    pages = "2240007",
    year = "2022"
}

@article{Olea:2005gb,
    author = "Olea, Rodrigo",
    title = "{Mass, angular momentum and thermodynamics in four-dimensional Kerr-AdS black holes}",
    eprint = "hep-th/0504233",
    archivePrefix = "arXiv",
    doi = "10.1088/1126-6708/2005/06/023",
    journal = "JHEP",
    volume = "06",
    pages = "023",
    year = "2005"
}

@inproceedings{Gibbons:2002du,
    author = "Gibbons, G.",
    title = "{Euclidean quantum gravity: The view from 2002}",
    booktitle = "{Workshop on Conference on the Future of Theoretical Physics and Cosmology in Honor of Steven Hawking's 60th Birthday}",
    pages = "351--372",
    month = "1",
    year = "2002"
}

@article{Hawking:1978jn,
    author = "Hawking, Stephen W.",
    title = "{Euclidean Quantum Gravity}",
    reportNumber = "PRINT-78-0745 (CAMBRIDGE)",
    journal = "NATO Sci. Ser. B",
    volume = "44",
    pages = "145",
    year = "1979"
}

@article{Kontsevich:2021dmb,
    author = "Kontsevich, Maxim and Segal, Graeme",
    title = "{Wick Rotation and the Positivity of Energy in Quantum Field Theory}",
    eprint = "2105.10161",
    archivePrefix = "arXiv",
    primaryClass = "hep-th",
    doi = "10.1093/qmath/haab027",
    journal = "Quart. J. Math. Oxford Ser.",
    volume = "72",
    number = "1-2",
    pages = "673--699",
    year = "2021"
}

@article{Witten:2021nzp,
    author = "Witten, Edward",
    title = "{A Note On Complex Spacetime Metrics}",
    eprint = "2111.06514",
    archivePrefix = "arXiv",
    primaryClass = "hep-th",
    month = "11",
    year = "2021"
}

@article{Visser:2021ucg,
    author = "Visser, Matt",
    title = "{Feynman\textquoteright{}s i\ensuremath{\epsilon} prescription, almost real spacetimes, and acceptable complex spacetimes}",
    eprint = "2111.14016",
    archivePrefix = "arXiv",
    primaryClass = "gr-qc",
    doi = "10.1007/JHEP08(2022)129",
    journal = "JHEP",
    volume = "08",
    pages = "129",
    year = "2022"
}

@article{Brown:2015bva,
    author = "Brown, Adam R. and Roberts, Daniel A. and Susskind, Leonard and Swingle, Brian and Zhao, Ying",
    title = "{Holographic Complexity Equals Bulk Action?}",
    eprint = "1509.07876",
    archivePrefix = "arXiv",
    primaryClass = "hep-th",
    doi = "10.1103/PhysRevLett.116.191301",
    journal = "Phys. Rev. Lett.",
    volume = "116",
    number = "19",
    pages = "191301",
    year = "2016"
}

@incollection{Hawking:1979ig,
    author = "Hawking, S. W.",
    title = "{The path-integral approach to quantum gravity}",
    booktitle = "{General Relativity: An Einstein Centenary Survey}",
    editor="Hawking, S. W. and Israel, W.",
    publisher = "Univ. Press",
    address="Cambridge, UK",
    month = "",
    year = "1979"
}

@article{Hawking:1978jz,
    author = "Hawking, S. W.",
    title = "{Quantum Gravity and Path Integrals}",
    doi = "10.1103/PhysRevD.18.1747",
    journal = "Phys. Rev. D",
    volume = "18",
    pages = "1747--1753",
    year = "1978"
}

@article{Adak:2023ymc,
    author = "Adak, Muzaffer and Dereli, Tekin and Koivisto, Tomi S. and Pala, Caglar",
    title = "{General teleparallel metrical geometries}",
    eprint = "2303.17812",
    archivePrefix = "arXiv",
    primaryClass = "gr-qc",
    doi = "10.1142/S0219887823502158",
    journal = "Int. J. Geom. Meth. Mod. Phys.",
    volume = "20",
    number = "Supp01",
    pages = "2350215",
    year = "2023"
}

@article{Cai:2015emx,
      author         = "Cai, Yi-Fu and Capozziello, Salvatore and De Laurentis,
                        Mariafelicia and Saridakis, Emmanuel N.",
      title          = "{f(T) teleparallel gravity and cosmology}",
      journal        = "Rept. Prog. Phys.",
      volume         = "79",
      year           = "2016",
      number         = "10",
      pages          = "106901",
      doi            = "10.1088/0034-4885/79/10/106901",
      eprint         = "1511.07586",
      archivePrefix  = "arXiv",
      primaryClass   = "gr-qc",
      SLACcitation   = "%%CITATION = ARXIV:1511.07586;%%"
}

@article{Cho:1975dh,
      author         = "Cho, Y. M.",
      title          = "{Einstein Lagrangian as the Translational Yang-Mills
                        Lagrangian}",
      journal        = "Phys. Rev.",
      volume         = "D14",
      year           = "1976",
      pages          = "2521",
      doi            = "10.1103/PhysRevD.14.2521",
      reportNumber   = "EFI 75-61-CHICAGO",
      SLACcitation   = "%%CITATION = PHRVA,D14,2521;%%"
}

@article{Ferraro:2006jd,
      author         = "Ferraro, Rafael and Fiorini, Franco",
      title          = "{Modified teleparallel gravity: Inflation without
                        inflaton}",
      journal        = "Phys. Rev.",
      volume         = "D75",
      year           = "2007",
      pages          = "084031",
      doi            = "10.1103/PhysRevD.75.084031",
      eprint         = "gr-qc/0610067",
      archivePrefix  = "arXiv",
      primaryClass   = "gr-qc",
      SLACcitation   = "%%CITATION = GR-QC/0610067;%%"
}

@article{Ferraro:2008ey,
      author         = "Ferraro, Rafael and Fiorini, Franco",
      title          = "{On Born-Infeld Gravity in Weitzenbock spacetime}",
      journal        = "Phys. Rev.",
      volume         = "D78",
      year           = "2008",
      pages          = "124019",
      doi            = "10.1103/PhysRevD.78.124019",
      eprint         = "0812.1981",
      archivePrefix  = "arXiv",
      primaryClass   = "gr-qc",
      SLACcitation   = "%%CITATION = ARXIV:0812.1981;%%"
}

@article{Linder:2010py,
      author         = "Linder, Eric V.",
      title          = "{Einstein's Other Gravity and the Acceleration of the
                        Universe}",
      journal        = "Phys. Rev.",
      volume         = "D81",
      year           = "2010",
      pages          = "127301",
      doi            = "10.1103/PhysRevD.81.127301, 10.1103/PhysRevD.82.109902",
      note           = "[Erratum: Phys. Rev.D82,109902(2010)]",
      eprint         = "1005.3039",
      archivePrefix  = "arXiv",
      primaryClass   = "astro-ph.CO",
      SLACcitation   = "%%CITATION = ARXIV:1005.3039;%%"
}

@article{Moller1961,
      author         = "M{\o}ller, Cristian",
      title          = "{Conservation Laws and Absolute Parallelism in General Relativity}",
      journal        = "K. Dan. Vidensk. Selsk. Mat. Fys. Skr. ",
      volume         = "1",
      year           = "1961",
      number         = "10",
      pages          = "1-50",
      
}

@article{Chen:2014qtl,
    author = "Chen, Pisin and Izumi, Keisuke and Nester, James M. and Ong, Yen Chin",
    title = "{Remnant Symmetry, Propagation and Evolution in $f$(T) Gravity}",
    eprint = "1412.8383",
    archivePrefix = "arXiv",
    primaryClass = "gr-qc",
    doi = "10.1103/PhysRevD.91.064003",
    journal = "Phys. Rev. D",
    volume = "91",
    number = "6",
    pages = "064003",
    year = "2015"
}

@article{Ong:2013qja,
      author         = "Ong, Yen Chin and Izumi, Keisuke and Nester, James M. and
                        Chen, Pisin",
      title          = "{Problems with Propagation and Time Evolution in f(T)
                        Gravity}",
      journal        = "Phys. Rev.",
      volume         = "D88",
      year           = "2013",
      pages          = "024019",
      doi            = "10.1103/PhysRevD.88.024019",
      eprint         = "1303.0993",
      archivePrefix  = "arXiv",
      primaryClass   = "gr-qc",
      SLACcitation   = "%%CITATION = ARXIV:1303.0993;%%"
}

@article{Ferraro:2014owa,
      author         = "Ferraro, Rafael and Fiorini, Franco",
      title          = "{Remnant group of local Lorentz transformations in
                        $\mathcal{f}(T)$ theories}",
      journal        = "Phys. Rev.",
      volume         = "D91",
      year           = "2015",
      number         = "6",
      pages          = "064019",
      doi            = "10.1103/PhysRevD.91.064019",
      eprint         = "1412.3424",
      archivePrefix  = "arXiv",
      primaryClass   = "gr-qc",
      SLACcitation   = "%%CITATION = ARXIV:1412.3424;%%"
}

@article{Obukhov:2006sk,
      author         = "Obukhov, Yuri N. and Rubilar, Guillermo F.",
      title          = "{Covariance properties and regularization of conserved currents in tetrad gravity}",
      journal        = "Phys. Rev.",
      volume         = "D73",
      year           = "2006",
      pages          = "124017",
      doi            = "10.1103/PhysRevD.73.124017",
      eprint         = "gr-qc/0605045",
      archivePrefix  = "arXiv",
      primaryClass   = "gr-qc",
      reportNumber   = "IFT-P-015-2006",
      SLACcitation   = "%%CITATION = GR-QC/0605045;%%"
}

@article{Golovnev:2020zpv,
    author = "Golovnev, Alexey and Guzm{\'a}n, Mar{\'\i}a-Jos{\'e}",
    title = "{Foundational issues in f(T) gravity theory}",
    eprint = "2012.14408",
    archivePrefix = "arXiv",
    primaryClass = "gr-qc",
    doi = "10.1142/S0219887821400077",
    journal = "Int. J. Geom. Meth. Mod. Phys.",
    volume = "18",
    number = "supp01",
    pages = "2140007",
    year = "2021"
}

@article{Maluf:2013gaa,
      author         = "Maluf, J. W.",
      title          = "{The teleparallel equivalent of general relativity}",
      journal        = "Annalen Phys.",
      volume         = "525",
      year           = "2013",
      pages          = "339-357",
      doi            = "10.1002/andp.201200272",
      eprint         = "1303.3897",
      archivePrefix  = "arXiv",
      primaryClass   = "gr-qc",
      SLACcitation   = "%%CITATION = ARXIV:1303.3897;%%"
}

@article{Fadeev,
      author         = "Faddeev, L. D.",
      title          = "{The energy problem in Einstein's theory of gravitation}",
      journal        = "Sov. Phys. Usp.",
      volume         = "25",
      year           = "1982",
      pages          = "130-142",
      doi            = "10.1070/PU1982v025n03ABEH004517",
}

@article{Maluf:1995re,
    author = "Maluf, Jose W.",
    title = "{Localization of energy in general relativity}",
    eprint = "gr-qc/9504010",
    archivePrefix = "arXiv",
    doi = "10.1063/1.530959",
    journal = "J. Math. Phys.",
    volume = "36",
    pages = "4242--4247",
    year = "1995"
}

@article{Maluf:1996kx,
    author = "Maluf, J. W. and Martins, E. F. and Kneip, A.",
    title = "{Gravitational energy of rotating black holes}",
    eprint = "gr-qc/9608049",
    archivePrefix = "arXiv",
    doi = "10.1063/1.531778",
    journal = "J. Math. Phys.",
    volume = "37",
    pages = "6302--6310",
    year = "1996"
}

@article{Witten:1998qj,
    author = "Witten, Edward",
    title = "{Anti de Sitter space and holography}",
    eprint = "hep-th/9802150",
    archivePrefix = "arXiv",
    reportNumber = "IASSNS-HEP-98-15",
    doi = "10.4310/ATMP.1998.v2.n2.a2",
    journal = "Adv. Theor. Math. Phys.",
    volume = "2",
    pages = "253--291",
    year = "1998"
}

@article{Brezina:2025dbc,
    author = "Brezina, Sebastian and Boffo, Eugenia and Kr{\v{s}}{\v{s}}{\'a}k, Martin",
    title = "{Teleparallel gravity from the principal bundle viewpoint}",
    eprint = "2507.23672",
    archivePrefix = "arXiv",
    primaryClass = "gr-qc",
    doi = "10.1007/JHEP05(2026)182",
    journal = "JHEP",
    volume = "05",
    pages = "182",
    year = "2026"
}

@article{Regge:1974zd,
    author = "Regge, Tullio and Teitelboim, Claudio",
    title = "{Role of Surface Integrals in the Hamiltonian Formulation of General Relativity}",
    reportNumber = "Print-74-0988 (IAS,PRINCETON)",
    doi = "10.1016/0003-4916(74)90404-7",
    journal = "Annals Phys.",
    volume = "88",
    pages = "286",
    year = "1974"
}

@article{Obukhov:2002tm,
      author         = "Obukhov, {\relax Yu}. N. and Pereira, J. G.",
      title          = "{Metric affine approach to teleparallel gravity}",
      journal        = "Phys. Rev.",
      volume         = "D67",
      year           = "2003",
      pages          = "044016",
      doi            = "10.1103/PhysRevD.67.044016",
      eprint         = "gr-qc/0212080",
      archivePrefix  = "arXiv",
      primaryClass   = "gr-qc",
      reportNumber   = "IFT-P-94-2002",
      SLACcitation   = "%%CITATION = GR-QC/0212080;%%"
}

@article{Lucas:2009nq,
      author         = "Lucas, Tiago Gribl and Obukhov, Yuri N. and Pereira, J.
                        G.",
      title          = "{Regularizing role of teleparallelism}",
      journal        = "Phys. Rev.",
      volume         = "D80",
      year           = "2009",
      pages          = "064043",
      doi            = "10.1103/PhysRevD.80.064043",
      eprint         = "0909.2418",
      archivePrefix  = "arXiv",
      primaryClass   = "gr-qc",
      SLACcitation   = "%%CITATION = ARXIV:0909.2418;%%"
}

@article{Krssak:2015lba,
    author = "Kr\v{s}\v{s}\'ak, Martin",
    title = "{Holographic Renormalization in Teleparallel Gravity}",
    eprint = "1510.06676",
    archivePrefix = "arXiv",
    primaryClass = "gr-qc",
    doi = "10.1140/epjc/s10052-017-4621-3",
    journal = "Eur. Phys. J. C",
    volume = "77",
    number = "1",
    pages = "44",
    year = "2017"
}

@article{Krssak:2015rqa,
      author         = "Kr\v{s}\v{s}\'ak, Martin and Pereira, J. G.",
      title          = "{Spin Connection and Renormalization of Teleparallel
                        Action}",
      journal        = "Eur. Phys. J.",
      volume         = "C75",
      year           = "2015",
      number         = "11",
      pages          = "519",
      doi            = "10.1140/epjc/s10052-015-3749-2",
      eprint         = "1504.07683",
      archivePrefix  = "arXiv",
      primaryClass   = "gr-qc",
      SLACcitation   = "%%CITATION = ARXIV:1504.07683;%%"
}

@article{Einstein1928b,
      author         = "Einstein, Albert",
      title          = "{Neue Möglichkeit für eine einheitliche Feldtheorie von Gravitation und Elektrizität}",
      journal        = "Sitzber. Preuss. Akad. Wiss. ",
      volume         = "17",
      year           = "1928",
      pages          = "224–227",
      
}

@article{Einstein1929a,
      author         = "Einstein, Albert",
      title          = "{Einheitliche Feldtheorie und Hamiltonsches Prinzip }",
      journal        = "Sitzber. Preuss. Akad. Wiss. ",
      volume         = "18",
      year           = "1929",
      pages          = "156–159",
      
}

@article{Maluf:2018coz,
    author = "Maluf, J.W. and Ulhoa, S.C. and da Rocha-Neto, J.F. and Carneiro, F.L.",
    title = "{Difficulties of Teleparallel Theories of Gravity with Local Lorentz Symmetry}",
    eprint = "1811.06876",
    archivePrefix = "arXiv",
    primaryClass = "gr-qc",
    doi = "10.1088/1361-6382/ab7288",
    journal = "Class. Quant. Grav.",
    volume = "37",
    number = "6",
    pages = "067003",
    year = "2020"
}

@article{Sauer:2004hj,
      author         = "Sauer, Tilman",
      title          = "{Field equations in teleparallel spacetime: Einstein's
                        fernparallelismus approach towards unified field theory}",
      journal        = "Historia Math.",
      volume         = "33",
      year           = "2006",
      pages          = "399-439",
      doi            = "10.1016/j.hm.2005.11.005",
      eprint         = "physics/0405142",
      archivePrefix  = "arXiv",
      primaryClass   = "physics",
      SLACcitation   = "%%CITATION = PHYSICS/0405142;%%"
}

@article{Balasubramanian:1999re,
    author = "Balasubramanian, Vijay and Kraus, Per",
    title = "{A Stress tensor for Anti-de Sitter gravity}",
    eprint = "hep-th/9902121",
    archivePrefix = "arXiv",
    reportNumber = "HUTP-99-A002, EFI-99-6, NSF-ITP-98-132",
    doi = "10.1007/s002200050764",
    journal = "Commun. Math. Phys.",
    volume = "208",
    pages = "413--428",
    year = "1999"
}

@article{Mann:2005yr,
    author = "Mann, Robert B. and Marolf, Donald",
    title = "{Holographic renormalization of asymptotically flat spacetimes}",
    eprint = "hep-th/0511096",
    archivePrefix = "arXiv",
    doi = "10.1088/0264-9381/23/9/010",
    journal = "Class. Quant. Grav.",
    volume = "23",
    pages = "2927--2950",
    year = "2006"
}

@article{Goenner:2004se,
    author = "Goenner, H. F. M.",
    title = "{On the history of unified field theories}",
    doi = "10.12942/lrr-2004-2",
    journal = "Living Rev. Rel.",
    volume = "7",
    pages = "2",
    year = "2004"
}

@article{Gibbons:1976ue,
      author         = "Gibbons, G. W. and Hawking, S. W.",
      title          = "{Action Integrals and Partition Functions in Quantum
                        Gravity}",
      journal        = "Phys. Rev.",
      volume         = "D15",
      year           = "1977",
      pages          = "2752-2756",
      doi            = "10.1103/PhysRevD.15.2752",
      reportNumber   = "PRINT-76-0995 (CAMBRIDGE)",
      SLACcitation   = "%%CITATION = PHRVA,D15,2752;%%"
}

@article{BeltranJimenez:2018vdo,
    author = "Beltr\'an Jim\'enez, Jose and Heisenberg, Lavinia and Koivisto, Tomi S.",
    title = "{Teleparallel Palatini theories}",
    eprint = "1803.10185",
    archivePrefix = "arXiv",
    primaryClass = "gr-qc",
    reportNumber = "NORDITA-2018-023, IFT-UAM/CSIC-18-035, IFT-UAM-CSIC-18-035",
    doi = "10.1088/1475-7516/2018/08/039",
    journal = "JCAP",
    volume = "08",
    pages = "039",
    year = "2018"
}

@article{Gomes:2023hyk,
    author = "Gomes, D\'ebora Aguiar and Beltr\'an Jim\'enez, Jose and Koivisto, Tomi S.",
    title = "{General parallel cosmology}",
    eprint = "2309.08554",
    archivePrefix = "arXiv",
    primaryClass = "gr-qc",
    doi = "10.1088/1475-7516/2023/12/010",
    journal = "JCAP",
    volume = "12",
    pages = "010",
    year = "2023"
}

@article{BeltranJimenez:2019bnx,
    author = "Beltr\'an Jim\'enez, Jose and Heisenberg, Lavinia and Koivisto, Tomi S.",
    title = "{The canonical frame of purified gravity}",
    eprint = "1903.12072",
    archivePrefix = "arXiv",
    primaryClass = "gr-qc",
    reportNumber = "NORDITA 2019-030",
    doi = "10.1142/S0218271819440127",
    journal = "Int. J. Mod. Phys. D",
    volume = "28",
    number = "14",
    pages = "1944012",
    year = "2019"
}

@article{Golovnev:2020nln,
    author = "Golovnev, Alexey and Guzman, Maria-Jose",
    title = "{Non-trivial Minkowski backgrounds in f(T) gravity}",
    eprint = "2012.00696",
    archivePrefix = "arXiv",
    primaryClass = "gr-qc",
    month = "12",
    year = "2020"
}

@article{Emtsova:2021ehh,
    author = "Emtsova, E. D. and Kr\v{s}\v{s}\'ak, M. and Petrov, A. N. and Toporensky, A. V.",
    title = "{On Conserved Quantities for the Schwarzschild Black Hole in Teleparallel Gravity}",
    eprint = "2105.13312",
    archivePrefix = "arXiv",
    primaryClass = "gr-qc",
    month = "5",
    year = "2021"
}

@book{Aldrovandi:2013wha,
    author = "Aldrovandi, Ruben and Pereira, Jos\'e Geraldo",
    title = "{Teleparallel Gravity}: {An Introduction}",
    doi = "10.1007/978-94-007-5143-9",
    isbn = "978-94-007-5142-2, 978-94-007-5143-9",
    publisher = "Springer",
    year = "2013"
}

@article{Samuel:2015oea,
    author = "Samuel, Joseph",
    title = "{Wick Rotation in the Tangent Space}",
    eprint = "1510.07365",
    archivePrefix = "arXiv",
    primaryClass = "gr-qc",
    doi = "10.1088/0264-9381/33/1/015006",
    journal = "Class. Quant. Grav.",
    volume = "33",
    number = "1",
    pages = "015006",
    year = "2016"
}

@article{Frob:2021dmv,
    author = {Fr\"ob, Markus B.},
    title = "{Kerr\textendash{}Schild metrics in teleparallel gravity}",
    eprint = "2103.02620",
    archivePrefix = "arXiv",
    primaryClass = "gr-qc",
    doi = "10.1140/epjc/s10052-021-09551-5",
    journal = "Eur. Phys. J. C",
    volume = "81",
    number = "8",
    pages = "766",
    year = "2021"
}

@article{Brown:2015lvg,
    author = "Brown, Adam R. and Roberts, Daniel A. and Susskind, Leonard and Swingle, Brian and Zhao, Ying",
    title = "{Complexity, action, and black holes}",
    eprint = "1512.04993",
    archivePrefix = "arXiv",
    primaryClass = "hep-th",
    doi = "10.1103/PhysRevD.93.086006",
    journal = "Phys. Rev. D",
    volume = "93",
    number = "8",
    pages = "086006",
    year = "2016"
}

@article{Stano:2025eje,
    author = "Stano, Michal",
    title = "{Black Hole Action in Einstein's other Gravity}",
    journal       = {Master's thesis, Department of Theoretical Physics, Comenius University in Bratislava},
    eprint = "2503.11746",
    archivePrefix = "arXiv",
    primaryClass = "gr-qc",
    year = 2024
}

@book{Nakahara:2003nw,
      author         = "Nakahara, M.",
      title          = "{Geometry, topology and physics}",
      publisher        = "Boca Raton, USA: Taylor \& Francis",
      year           = "2003",
      SLACcitation   = "%%CITATION = INSPIRE-640779;%%"
}

@article{Visser:2017atf,
    author = "Visser, Matt",
    title = "{How to Wick rotate generic curved spacetime}",
    eprint = "1702.05572",
    archivePrefix = "arXiv",
    primaryClass = "gr-qc",
    month = "2",
    year = "2017"
}

@article{Gomes:2022vrc,
    author = "Gomes, D{\'e}bora Aguiar and Beltr{\'a}n Jim{\'e}nez, Jose and Koivisto, Tomi S.",
    title = "{Energy and entropy in the geometrical trinity of gravity}",
    eprint = "2205.09716",
    archivePrefix = "arXiv",
    primaryClass = "gr-qc",
    doi = "10.1103/PhysRevD.107.024044",
    journal = "Phys. Rev. D",
    volume = "107",
    number = "2",
    pages = "024044",
    year = "2023"
}

@article{Emparan:1999pm,
    author = "Emparan, Roberto and Johnson, Clifford V. and Myers, Robert C.",
    title = "{Surface terms as counterterms in the AdS / CFT correspondence}",
    eprint = "hep-th/9903238",
    archivePrefix = "arXiv",
    reportNumber = "DTP-99-21, UK-99-04, MCGILL-99-12, EHU-FT-9906",
    doi = "10.1103/PhysRevD.60.104001",
    journal = "Phys. Rev. D",
    volume = "60",
    pages = "104001",
    year = "1999"
}

@article{Anastasiou:2020zwc,
    author = "Anastasiou, Giorgos and Miskovic, Olivera and Olea, Rodrigo and Papadimitriou, Ioannis",
    title = "{Counterterms, Kounterterms, and the variational problem in AdS gravity}",
    eprint = "2003.06425",
    archivePrefix = "arXiv",
    primaryClass = "hep-th",
    reportNumber = "KIAS-P20010",
    doi = "10.1007/JHEP08(2020)061",
    journal = "JHEP",
    volume = "08",
    pages = "061",
    year = "2020"
}

@article{Henningson:1998gx,
    author = "Henningson, M. and Skenderis, K.",
    title = "{The Holographic Weyl anomaly}",
    eprint = "hep-th/9806087",
    archivePrefix = "arXiv",
    reportNumber = "CERN-TH-98-188, KUL-TF-98-21",
    doi = "10.1088/1126-6708/1998/07/023",
    journal = "JHEP",
    volume = "07",
    pages = "023",
    year = "1998"
}

@article{Bianchi:2001kw,
    author = "Bianchi, Massimo and Freedman, Daniel Z. and Skenderis, Kostas",
    title = "{Holographic renormalization}",
    eprint = "hep-th/0112119",
    archivePrefix = "arXiv",
    reportNumber = "MIT-CTP-3166, PUTP-1999, DAMTP-2001-63, ROM2F-2001-30",
    doi = "10.1016/S0550-3213(02)00179-7",
    journal = "Nucl. Phys. B",
    volume = "631",
    pages = "159--194",
    year = "2002"
}

@article{Skenderis:2002wp,
    author = "Skenderis, Kostas",
    editor = "de Wit, B. and Vandoren, S.",
    title = "{Lecture notes on holographic renormalization}",
    eprint = "hep-th/0209067",
    archivePrefix = "arXiv",
    reportNumber = "PUTP-2047",
    doi = "10.1088/0264-9381/19/22/306",
    journal = "Class. Quant. Grav.",
    volume = "19",
    pages = "5849--5876",
    year = "2002"
}
\bibliographystyle{Style}
\end{document}